\documentstyle[preprint,aps,epsfig]{revtex}
\begin{document}
\tighten
\title{ Microscopic model analyses of the elastic scattering 
of 65 MeV protons from targets of diverse mass}
\author{P. J. Dortmans and K. Amos}
\address{ School of Physics, University of Melbourne,
Parkville 3052, Victoria, Australia}
\author{S. Karataglidis}
\address{ TRIUMF, 4004 Wesbrook Mall, Vancouver,
British Columbia, V6T 2A3, Canada }
\author{J. Raynal}
\address{ C.E.A.-Saclay, Service de Physique Th\'{e}orique, 
F-91191 Gif-sur-Yvette Cedex, France}

\date{\today}
\maketitle

\begin{abstract}
Nonlocal coordinate space optical potentials for the scattering of 65
MeV protons from nuclei ranging in mass from $^6$Li to $^{238}$U have
been defined by folding a complex, medium dependent effective
interaction with the density matrix elements of each target.  The
effective interaction is based upon solutions of the
Lippmann--Schwinger and Brueckner--Bethe--Goldstone equations having
the Paris potential as input. The nuclear structure information
required in our folding model are the one body density matrix elements
for the target and the single nucleon bound state wave functions that
they weight. For light mass nuclei, very large basis shell model
calculations have been made to obtain the one body density matrix
elements.  For medium and heavy mass nuclei, a very simple shell model
prescription has been used.  The bound state single particle wave
functions that complete the nuclear density matrices are either
Woods--Saxon or harmonic oscillator functions.  The former are
employed in most cases when large basis structure is available.  For
light nuclei ($A\le 16$) Woods--Saxon potential parameters and
harmonic oscillator lengths are determined from fits to electron
scattering form factors.  For all other nuclei, we use harmonic
oscillator functions with the oscillator lengths set from an $A^{1/6}$
mass law. Using this microscopic model, optical potentials result from
which differential cross sections, analyzing powers and spin rotations
are found. In general the calculated results compare very well with
data when the effective interactions are determined from a mapping of
nucleon--nucleon $g$ matrices. This is not the case when effective
interactions built from a mapping of (free) $t$ matrices are used.
\end{abstract}


\section{Introduction}

A microscopic, coordinate space model to analyse proton scattering
from nuclei has been developed~\cite{Dort94,C200} from that formed
earlier by the Hamburg group~\cite{Rikus}.  With this model, analyses
have been made of 200 MeV proton elastic scattering from a large
number of nuclei~\cite{mass200}, of elastic and inelastic p--$^{12}$C
scattering over a range of incident energies~\cite{C12energy}, of
charge exchange reactions~\cite{CEreaction}, of the structure of
neutron rich nuclei such as $^{6,8}$He~\cite{Varenna} and
$^{9,11}$Li~\cite{Li911ref}, and, very recently, of proton scattering
from ${}^{3,4}$He~\cite{He3paper}. Those light mass systems usually
have been considered as few--body problems and the data analysed with
few--body methods.  We note also that microscopic model analyses
determined within a momentum space framework for elastic proton
scattering have been made, with varying degrees of
success~\cite{Elst90,Arel95,Arel96,Elst98}.  The energies considered
by these microscopic analyses (both in coordinate and momentum space)
lie in a `transition' region between low and intermediate energies.
For such energies, effects of nonlocalities in the effective
nucleon--nucleus ($NA$) interaction must be taken into account.  Also
for these energies, medium dependent effects in the nucleon--nucleon
($NN$) effective interaction upon which that $NA$ interaction is
built, are important.  When these two facets are taken into account in
our coordinate space approach, good to excellent agreement with data
is found to measured elastic and inelastic scattering cross sections,
analyzing powers and other spin measureables~\cite{mass200,Arel95}.
The optical potentials are formed by folding the complex effective
interaction with nuclear one body density matrix elements (OBDME) and
a set of single particle (SP) bound state functions. For very light
nuclei, modern shell model studies~\cite{Barrett} not only specify the
OBDME but also which SP wave functions should be used. Using this
information, calculations of both electron scattering form factors and
proton elastic scattering observables become predictive; the latter
especially when the effective interaction and folding procedure are
fixed. Otherwise, one can use fits to the longitudinal elastic
electron scattering form factor to specify the SP wave functions one
should use in the $NA$ scattering calculations.  Those $NA$
calculations then remain predictive as there is no adjustable quantity
left.  For heavier nuclei (A $>$ 16), shell model calculations to date
have been made only within $0\hbar\omega$ space.  Therefore we have
not used electron scattering form factor analyses to select a
definitive set of SP wave functions. However, experience
suggests~\cite{mass200} that harmonic oscillator (HO) functions with
oscillator lengths following a simple mass rule ($b=A^{1/6}$) should
suffice.

Whatever the choice of structure input, the folding process leads to
$NA$ optical potentials that are nonlocal because of exchange
(antisymmetry) contributions.  In our model, antisymmetrization of the
proton--target system has been taken at the two--body level only,
i.e. we have used a fully antisymmetrized $A$--nucleon target wave
function and antisymmetrized each projectile--target nucleon pair.  In
the past, the resultant nonlocality of the optical potential either
was ignored or was approximated by an equivalent local
form~\cite{Rikus}.  Our calculations of 200 MeV proton--nucleus ($pA$)
scattering gave excellent predictions of observables but only when the
complete integro--differential forms of Schr\"{o}dinger equations were
used~\cite{mass200}.

As with our study at 200 MeV, we consider herein only the elastic
scattering channel but take the spin rotation, $R$, into account along
with the cross sections, $d\sigma /d\Omega$, and analyzing powers,
$A_y$.  Specifically we have considered 50 targets, namely the
$0p$--shell nuclei $^{6,7,9,11}$Li, $^{11}$B, $^{12}$C, and $^{16}$O;
the $1s 0d$--shell nuclei $^{20}$Ne, $^{24}$Mg, $^{28}$Si, and
$^{32}$S; the $1s 0d$ proton--$0f 1p$ neutron shell nucleus,
$^{40}$Ar; the $0f 0p$--shell $^{40,42,44,48}$Ca, $^{46,48,50}$Ti,
$^{52}$Cr, $^{54,56}$Fe, $^{59}$Co, and $^{58,60,62,64}$Ni; the $2s 1d
0g$ neutron shell nuclei $^{89}$Y, $^{90}$Zr, $^{98,100}$Mo, and
$^{118}$Sn; the lanthanide nuclei, $^{144,152,154}$Sm, $^{160}$Gd,
$^{164}$Dy, $^{166,168}$Er and $^{174,176}$Yb; the $0h_{11/2}$
proton--$0i_{13/2}$ neutron shell nuclei $^{178,180}$Hf,
$^{182,184}$W, and $^{192}$Os; the mass 200 nuclei, $^{208}$Pb and
$^{209}$Bi; and the actinide pair, $^{232}$Th and $^{238}$U.  Our
predictions of elastic scattering from these diverse mass targets are
compared with 65 MeV data in all cases with the exception of
$^{9,11}$Li; the experimental data for which come from the elastic
scattering of radioactive beams of those lithium isotopes from
hydrogen.  Inverse kinematics makes those experiments equivalent to 60
and 62 MeV protons scattering from $^{9}$Li and $^{11}$Li,
respectively.

We compare predictions obtained from the optical potentials defined
for each target with the proton elastic scattering data that is
available, but only to 80$^\circ$ in the centre of mass. In general,
the cross--section magnitudes by that scattering angle are typically
$\sim 1$~mb/sr. We do not expect the approximations needed to make our
model practical would be appropriate necessarily in making a
prediction of the scattering of lesser magnitude.

The paper is arranged as follows. The procedure for obtaining our
microscopic optical potentials is outlined in Section~\ref{sec_opt},
as are the amplitudes by which we obtain the proton scattering
observables. In Section~\ref{sec_results} we present and discuss the
results for the scattering of 65 MeV protons from those nuclei
included in the present study. Conclusions are drawn in
Section~\ref{sec_conc}.


\section{The microscopic optical potential}
\label{sec_opt}
We develop first the folding procedure by which the nonlocal optical
potentials are specified. From those, the effective $NA$ interaction
is obtained and we define the amplitudes that specify the scattering
observables.

\subsection{The folding process}

In a representation with ${\bf r, r^\prime}$ denoting relative
coordinates between a colliding pair of particles, the Schr\"odinger
equation describing their scattering by a local Coulomb, $V_C(r)$,
plus a nonlocal hadronic (optical) potential takes the form
\begin{equation}
\left[ \frac{\hbar^2}{2\mu} \nabla^2 - V_C(r) + E \right] \Psi({\bf
r}) = \int U({\bf r, r^\prime}) \Psi({\bf r^\prime})\; d{\bf r^\prime} \;.
\label{eq:nls}
\end{equation}
This reduces by using the partial wave expansions, 
\begin{equation}
\Psi({\bf r}) = \sum_{lm} \frac{u_l(r)}{r} i^l Y_{lm}(\Omega_r)\;,
\end{equation}
and
\begin{equation}
U({\bf r},{\bf r^\prime})
= \sum_{lm} \frac{ W_l(r,r') }{ rr' } i^l Y_{lm}(\Omega_r) i^{-l}
Y^{\ast}_{lm}(\Omega_{r'}) \;,
\label{eq:Wexpand}
\end{equation}
to a set of integro--differential equations
\begin{equation}
\left\{ \frac{\hbar^2}{2\mu} \left[ \frac{d^2}{dr^2} -
\frac{l(l+1)}{r^2} \right] - V_C(r) + E \right\} u_l(r) =
\int_0^\infty W_l(r,r') u_l(r') \;dr' \;.
\label{eq:nldinger}
\end{equation}
The $W_{l}(r,r^\prime)$ contain both the local interaction and
multipoles of the nonlocal interaction. Note that for simplicity, we
have suppressed all terms due to the intrinsic spin of the system. We
seek solutions for $NA$ scattering and determine the nonlocal
interactions, $U_{NA}({\bf r},{\bf r^\prime})$, at 65 MeV in
particular, by folding effective $NA$ interactions with the relevant
structure information. Thus we obtain the appropriate $NA$ effective
interaction for each nucleus in our investigation from $^6$Li to
$^{238}$U.  At this particular energy one may anticipate stronger
influences in analyses of the medium effects defining the effective
$NN$ interactions and of the nonlocalities in the optical potentials
arising from the folding process than in studies of 200 MeV
scattering~\cite{mass200}. With the optical potential in this
coordinate space representation, we use the program
DWBA91~\cite{Raynal} to solve the set of partial wave
integro--differential Schr\"{o}dinger equations.  That code has the
further useful attribute that it can be used to evaluate distorted
wave born approximated (DWBA) amplitudes for inelastic scattering,
given the appropriate OBDME and SP wave functions. The microscopic
optical potentials are used therein to determine the distorted waves
and the medium modified effective $NN$ interaction is used as the
transition operator. Data from inelastic scattering of 200 MeV protons
from $^{6,7}$Li and $^{12}$C have been well reproduced by using that
procedure~\cite{C200,Li67} further justifying the scattering theory
formulated in coordinate space.

To define the nonlocal interaction for $NA$ scattering in a full
folding model, we must accommodate antisymmetry between the projectile
and every nucleon specified with the internal nuclear wave function.
We must evaluate multi--particle matrix elements of the form
\begin{equation}
U_{pA} = \left\langle \Psi \left( 1 \ldots A \right) \left|
\sum_{N=1}^A V_{N0} \right| \Psi \left( 1 \ldots A \right) \right\rangle
\end{equation}
with $ \left\langle {\bf R} \right| \left. \Psi \left( 1 \ldots A
\right) \right\rangle $ being the many--body wave function for the
ground state of the target and `0' denoting the projectile
coordinates.  As all nucleons in the target are equivalent, it is
useful to choose a specific entry (``1'') and write
\begin{equation}
U_{pA} = A\left\langle \Psi \left( 1 \ldots A \right) \right|
V_{10} \left| \Psi \left( 1 \ldots A \right) \right\rangle\;.
\label{eq:U_multi}
\end{equation}
With the many--body state expanded in cofactors, 
\begin{equation}
\left| \Psi \left( 1 \ldots A \right) \right\rangle =
\frac{1}{\sqrt{A}} \sum_{\alpha m} \left| \varphi_{\alpha m}(1)
\right\rangle a_{\alpha m} \left| \Psi \left( 1 \ldots A \right) \right\rangle
\end{equation}
where $\alpha$ specifies the set $\{n,l,j,\zeta \}$, and $\zeta$ is
the isospin projection, Eq.~(\ref{eq:U_multi}) becomes
\begin{equation}
U_{pA} = \sum_{\alpha m \alpha' m'} \left\langle \Psi \left|
a^{\dagger}_{\alpha' m'} a_{\alpha m} \right| \Psi \right\rangle
\left\langle \varphi_{\alpha' m'}(1) \right| V_{10} \left\{ \left|
\varphi_{\alpha m}(1) \right\rangle - \left| \varphi_{\alpha m}(0)
\right\rangle \right\} \;,
\end{equation}
when the required antisymmetry with projectile and struck nucleon is
taken into account.  The nuclear structure information required to
evaluate the optical potentials are many--body matrix elements of the
particle--hole operators. They are defined by
\begin{eqnarray}
\rho_{\alpha \alpha' J_i J_f}^{m m' M_i M_f} & = & \left\langle \Psi \left|
a^{\dagger}_{\alpha' m'} a_{\alpha m} \right| \Psi \right\rangle
\nonumber \\
& = & \sum_{I,N} \frac{(-1)^{j - m}}{\sqrt{2J_f + 1}} \left\langle
\left. j \, m \, j' \, -m' \right| I \, -N \right\rangle
\nonumber \\
& \times & \left\langle  \left. J_i \, M_i \, I \, N \right| J_f \,
M_f \right\rangle S_{\alpha \alpha' I}
\end{eqnarray}
where the OBDME, $S_{\alpha \alpha' I}$, are 
(with  $\tilde{a}_{\alpha m} = (-1)^{j-m} a_{\alpha -m}$)
\begin{eqnarray}
S_{\alpha \alpha' I} & = &
\left\langle \Psi_{J_f} \left\| \left[ a_{\alpha'}^{\dagger} \times
\tilde{a}_{\alpha}
 \right]^I \right\| \Psi_{J_i} \right\rangle \nonumber \\
& \rightarrow & \left\langle \Psi_J \left\| \left[ a^{\dagger}_{\alpha'}
\times \tilde{a}_{\alpha} \right]^I \right\| \Psi_J \right\rangle \;,
\label{obdme}
\end{eqnarray}
in the elastic scattering case (from a target with spin $J$).  If $J$
is nonzero, multipoles from 0 to $I_{max} (= 2J)$
contribute. Scattering from light odd mass targets show that these
must be included~\cite{Li67}.  As even--even nuclei all have ground
state spin--parities ($J^\pi$) of $0^+$, the required OBDME simply are
the monopoles
\begin{equation}
S_{\alpha \alpha' 0} =
\left\langle \Psi_{0} \left\| \left[ a_{\alpha'}^{\dagger}\times
\tilde{a}_{\alpha} \right]^0 \right\| \Psi_{0} \right\rangle \;.
\end{equation}
While the angular momentum selection rules require $l = l'$ and $j =
j'$, there is no such restriction on the principle quantum number. The
cases where $n \neq n'$ signify purely radial excitations which can
only occur in spaces allowing for cross shell transitions. In those
instances, the specification of the full density requires the
inclusion of those off-diagonal elements. The diagonal elements reduce
to the shell occupancies as
\begin{equation}
\left\langle \Psi 
\left| a^{\dagger}_{\alpha' m'} a_{\alpha m} \right|
\Psi \right\rangle
\rightarrow 
\delta_{\alpha \alpha'} \delta_{m m'} 
\sigma_{\alpha \alpha'} \;,
\end{equation}
where $\sigma_{\alpha \alpha}$ is the fractional shell occupancy as a
fully occupied shell has $\sigma_{\alpha \alpha} = 1$.  Thus the
(diagonal) OBDME are given by
\begin{equation}
S_{\alpha \alpha 0} = \sqrt{2j + 1} \ 
\sigma_{\alpha \alpha} \;.
\end{equation}

The optical potential given by this folding process takes the form
\begin{eqnarray}
U({\bf r_1},{\bf r_2};E) & = & \sum_{\alpha m \alpha' m'}
\sigma_{\alpha \alpha'} \left[ \delta({\bf r_1}-{\bf r_2}) \int
\varphi^\ast_{\alpha' m'}({\bf s}) U^{D}(R_{1s},E) \varphi_{\alpha
m}({\bf s})\; d{\bf s} \right. \nonumber \\
& \phantom{ = }&\hspace*{3.0cm} + \left.  \varphi^\ast_{\alpha' m'}({\bf r_1})
U^{Ex}(R_{12},E) \varphi_{\alpha m}({\bf r_2}) \right] \; ,
\label{localnon}
\end{eqnarray}
where $R_{12} = |{\bf r_1}-{\bf r_2}|$, and $U^{D}$ and $U^{Ex}$ are
appropriate combinations of the multipoles of the effective
interaction for the direct and exchange contributions to the optical
potential respectively~\cite{Rikus,Raynal}.  One then applies the
partial wave expansions, Eq.~(\ref{eq:Wexpand}).

A feature in our process of analysis is that all details required to
make the calculations are preset. In the cases of
$^{6,7}$Li~\cite{Li67} and of $^{12}$C~\cite{C200}, nuclear structure
information was taken from shell model calculations in complete
multi-$\hbar\omega$ spaces with the SP wave functions set by fits to
the longitudinal elastic electron scattering form factors.  Thus the
results for both the elastic and inelastic scattering of 200 MeV
protons from those nuclei were obtained from single calculations.  No
adjustments (such as those attributed to core polarizations) were
needed and most proton excitation data were well described.  With the
microscopic (coordinate space) model for 200 MeV $pA$ scattering
established for a set of $0p$--shell nuclei, an extensive analysis for
many more nuclei with masses up to $^{208}$Pb was
completed~\cite{mass200}. In almost all cases single calculations led
to excellent reproduction of data.  Most recently~\cite{He3paper},
wave functions from very large space shell model calculations of
$^3$He were used in successful predictions of the elastic scattering
of 200 MeV protons from that few nucleon system. In that case the
shell model interaction used also fixed the SP wave functions so the
analysis of the electron scattering form factor was also predictive.

\subsection{The effective interaction}

The folding procedure to define the optical potential requires the
specification of the $NN$ effective interaction in $ST$-channel form
and in coordinate space.  This effective interaction we take as a
mixture of central, two--body spin--orbit and tensor force attributes,
each having a form factor that is a set of four Yukawa functions with
complex coefficients, i.e.
\begin{equation}
g_{\em eff}^{(i)ST)}(r,E) = \sum_{k=1}^{n_i} S_k^{(i)} \left[
\rho(r),E \right] \frac{ e^{-(\mu_k^{(i)} r)} }{r} \;.
\end{equation} 
Therein $S^{(i)}_{k} \left[ \rho(r),E \right]$ are complex strengths
that vary with projectile energy and nuclear density, $\mu^{(i)}_{k}$
are the inverse ranges of the Yukawa functions with $k$ the index for
those inverse ranges. In principle, the number of strengths and
inverse ranges ($n_i$) chosen can be as large as one likes, though for
all operators $n_{i} = 4$ seems to be sufficient to reproduce
accurately the half--off--shell $g$ matrices for laboratory energies
between 50 and 400 MeV~\cite{Dort94}. Singular valued decomposition
was used to optimise those inverse ranges and coefficients so that
double Bessel transforms of the effective interaction map accurately
to an appropriate set of infinite nuclear matter $g$
matrices~\cite{Haft70,Dort91} obtained from solution of the
Bethe--Brueckner--Goldstone ($BBG$) equation,
\begin{eqnarray}
g_{LL'}^{(JST)}(p',p;k,K,k_F) & = & V_{LL'}^{(JST)}(p,p') \nonumber \\
& + & \frac{2}{\pi} \sum_l \int_0^{\infty} V_{Ll}^{(JST)}(p',q) \;
[{\cal H}]\; g_{lL'}^{(JST)}(q,p;k,K,k_F) \; q^2 dq,
\end{eqnarray}
where
\begin{equation}
{\cal H}(q,k,K,k_f) =
\frac{\bar{Q}(q,K,k_f)}{\bar{E}(k,K,k_f) - \bar{E}(q,K,k_f) + i\varepsilon}
\end{equation}
in which $\bar{Q}(q,K,k_f)$ is an angle averaged Pauli operator with
an average center of mass momentum, $K$~\cite{Haft70,Dort91}.  The
energies in the propagators of the $BBG$ equations include auxiliary
potentials, $U$, and are defined by
\begin{equation}
\bar E(q,K,k_f)= \left( q^2 + K^2 \right) +
\left( \frac{m}{\hbar^2} \right)
\left\{ U\left( \left| {\bf q + K } \right| \right) +
U\left( \left| {\bf q - K } \right| \right) \right\}\ ,
\label{Ebar}
\end{equation}
wherein $U$ is the first order mass operator term.  We take
$V_{LL'}^{(JST)}(p,p')$ to be the Paris $NN$ interaction \cite{Paris},
although there is little difference if one starts with the Bonn $NN$
potential \cite{C200}. Details of the techniques involved are given
elsewhere~\cite{Haft70,Dort91}.

Given that the $NN$ $g$ matrices are most easily specified in momentum
space and the effective interaction form is an approximation, it is
sensible to seek to analyse $NA$ elastic scattering with a momentum
space solution of the Schr\"{o}dinger equation.  Such studies have
been made using credible $g$ matrices as input~\cite{Arel95,Arel96}
and those results reflect reasonable agreement with the data. They
also confirm the need for inclusion of medium effects for low and
intermediate energy $NA$ scattering.  Of note from the momentum space
calculations is the observation that off--shell Kowalski--Noyes
$f$--ratios of the $t$ and $g$ matrices vary quite
similarly~\cite{Dort91}. The major effect of medium modifications
(Pauli blocking and energy denominators) are to vary the on--shell
values.

The choice of the Yukawa forms for the effective interaction is
extremely advantageous when it comes to evaluating the optical
potential. Fourier transformation of each of the radial components of
the effective $NN$ interactions gives a simple multipole form, so that
for the central terms, the double Bessel transformation that leads to
each term in the nonlocal interaction can be solved analytically,
taking the form
\begin{equation}
W^{(k)}_{l}(r_1,r_2) \propto h^{(+)}_l\left( i\mu^{(k)} r_{>} \right) 
j_l\left( i\mu^{(k)} r_{<} \right),
\end{equation}
where $r_{<}$ and $r_{>}$ are the lesser and greater of $r_1$ and $r_2$
respectively.

\subsection{The observables}

There are diverse observables for the scattering of polarized protons
from an unpolarised target. While one may define differing sets, we
consider that which involves differential cross section, $d\sigma
/d\Omega$, analyzing power, $A_y$, and two Wolfenstein spin
rotations, $A$ and $R$. These measureables are defined in terms of
scattering amplitudes $F(\theta)$~\cite{Rayn72}. Since the data we
investigate were obtained using polarized projectile protons, this
amplitude is a $2\times 2$ matrix,
\begin{equation}
F(\theta) = 
\left|
\begin{array}{cc}
f_{\frac{1}{2} \frac{1}{2}}(\theta) & f_{\frac{1}{2} -\frac{1}{2}}(\theta) \\
-f_{\frac{1}{2} -\frac{1}{2}}(\theta) & f_{\frac{1}{2} \frac{1}{2}}(\theta) 
\end{array} \right|
\label{fdef}
\end{equation}
where
\begin{equation}
f_{{\frac{1}{2}}{\frac{1}{2}}}(\theta) 
= f_C(\theta) r^{\left( \frac{1}{2}
\right)}_{\frac{1}{2}\frac{1}{2}}(\theta)
+ \frac{1}{4ik} \sum_{J} (2J+1) e^{-2i\sigma^{C}_{J}}
\left( S^{+}_{J} + S^{-}_{J} -
2 \right) r^{(J)}_{\frac{1}{2} \frac{1}{2}}(\theta),
\label{adef}
\end{equation}
and
\begin{equation}
f_{\frac{1}{2} -\frac{1}{2}}(\theta) 
= f_C(\theta) r^{\left( \frac{1}{2}
\right)}_{\frac{1}{2} -\frac{1}{2}}(\theta) 
+ \frac{1}{4ik}\sum_{J} (2J+1) e^{-2i\sigma^{C}_{J}}
\left( S^{+}_{J} - S^{-}_{J} \right)
r^{(J)}_{\frac{1}{2} -\frac{1}{2}}(\theta).
\label{bdef}
\end{equation}
Here $r^{(J)}_{\frac{1}{2} \pm\frac{1}{2}}(\theta)$ are the rotation
matrices, $\sigma^{C}_{J}$ are the Coulomb phase shifts, $f_C(\theta)$
is the point Coulomb scattering amplitude, and $S^{\pm}_{J}$ are the
$S$ matrices for $J = L \pm \frac{1}{2}$. With these, the elastic
scattering observables are defined as
\begin{equation}
\frac{d\sigma}{d\Omega}(\theta) 
= \left| f_{\frac{1}{2} \frac{1}{2}}(\theta) \right|^2 
+ \left| f_{\frac{1}{2} -\frac{1}{2}}(\theta) \right|^2,
\label{csdef}
\end{equation}
\begin{equation}
A_y(\theta) = P(\theta) 
= \frac{2 \mbox{Im} \left\{ f_{\frac{1}{2} \frac{1}{2}}(\theta)
f_{\frac{1}{2} -\frac{1}{2}}^{\ast}(\theta) \right\} }%
{ \left| f_{\frac{1}{2} \frac{1}{2}}(\theta) \right|^2 
+ \left| f_{\frac{1}{2} -\frac{1}{2}}(\theta) \right|^2},
\label{Aydef}
\end{equation}
\begin{equation}
R(\theta) = K^x_x(\theta) = K^z_z(\theta)
= \frac{ \left| f_{\frac{1}{2} \frac{1}{2}}(\theta) \right|^2 
- \left| f_{\frac{1}{2} -\frac{1}{2}}(\theta) \right|^2}%
{ \left| f_{\frac{1}{2} \frac{1}{2}}(\theta) \right|^2 
+ \left| f_{\frac{1}{2} -\frac{1}{2}}(\theta) \right|^2 },
\label{WolfR}
\end{equation}
and
\begin{equation}
A(\theta) = K^x_z(\theta) = - K^z_x(\theta)
= \frac{ 2 \mbox{Re} \left\{ f_{\frac{1}{2} \frac{1}{2}}(\theta)
f_{\frac{1}{2} -\frac{1}{2}}^{\ast}(\theta) \right\} }%
{ \left| f_{\frac{1}{2} \frac{1}{2}}(\theta) \right|^2 
+ \left| f_{\frac{1}{2} -\frac{1}{2}}(\theta) \right|^2}.
\label{WolfA}
\end{equation}
Commonly $Q$, a linear combination of the Wolfenstein spin rotations,
is measured. It relates to the above by
\begin{equation}
Q(\theta) = R(\theta) \sin{\theta} + A(\theta) \cos{\theta}.
\label{SpinQ}
\end{equation}

\section{Results}
\label{sec_results}
In the following subsections, we display the results of our
calculations of the elastic scattering of 65 MeV protons from many
target nuclei and place them in comparison with experimental data
wherever such data exist.  The results are subdivided into four
sections.  First, we present the scattering from the light nuclei,
$^{6}$Li to $^{16}$O, for which OBDME have been obtained mostly from
large space shell model calculations and for which SP wave functions
have been determined from fits to the longitudinal elastic electron
scattering form factors. Then we present and discuss the results
obtained with medium mass nuclei, from $^{20}$Ne to $^{64}$Ni. For
those nuclei the shell model calculations were performed in complete
$0\hbar\omega$ spaces, except for those in the middle of the $fp$
shell where the dimension of the basis becomes prohibitively large for
matrix diagonalization techniques. In those cases, a restriction on
the number of nucleons in the $0f_{\frac{7}{2}}$ orbit was placed on
the model. The oscillator length for the HO SP wave functions used was
set as $A^{1/6}$ (in units of fm).  In the third section, we present
the results for the scattering from heavy nuclei, and in which the
nucleon occupancies are those given by a simple packing model. Again
we use HO wave functions with an oscillator length given by
$A^{1/6}$~fm as the SP functions. In the final section, we consider
the changing structure of the observables across the mass range.

For each of the cases discussed, we present the results of two
calculations. They differ only in the specification of the optical
potential. The results obtained with the effective interaction built
from the $NN$ $g$ matrices folded with known density
profiles~\cite{DeJa74} are displayed by the solid curves while those
found by using the free $NN$ $t$ matrices are displayed by the dashed
curves. For simplicity we designate the results as being obtained from
calculations made using $g$--folding and $t$--folding potentials
respectively. The data come from
Refs.~\cite{Tosa85,Moon92,Kuwa90,Yoso85,Kato85,Saka82,Noro81,Ohta83,Ichi87,Ichi84,Ogaw86,Take86}
with specific reference made in the figure captions.

\subsection{Light mass nuclei (A $\leq$ 16)}

\begin{figure}
\centering\epsfig{file=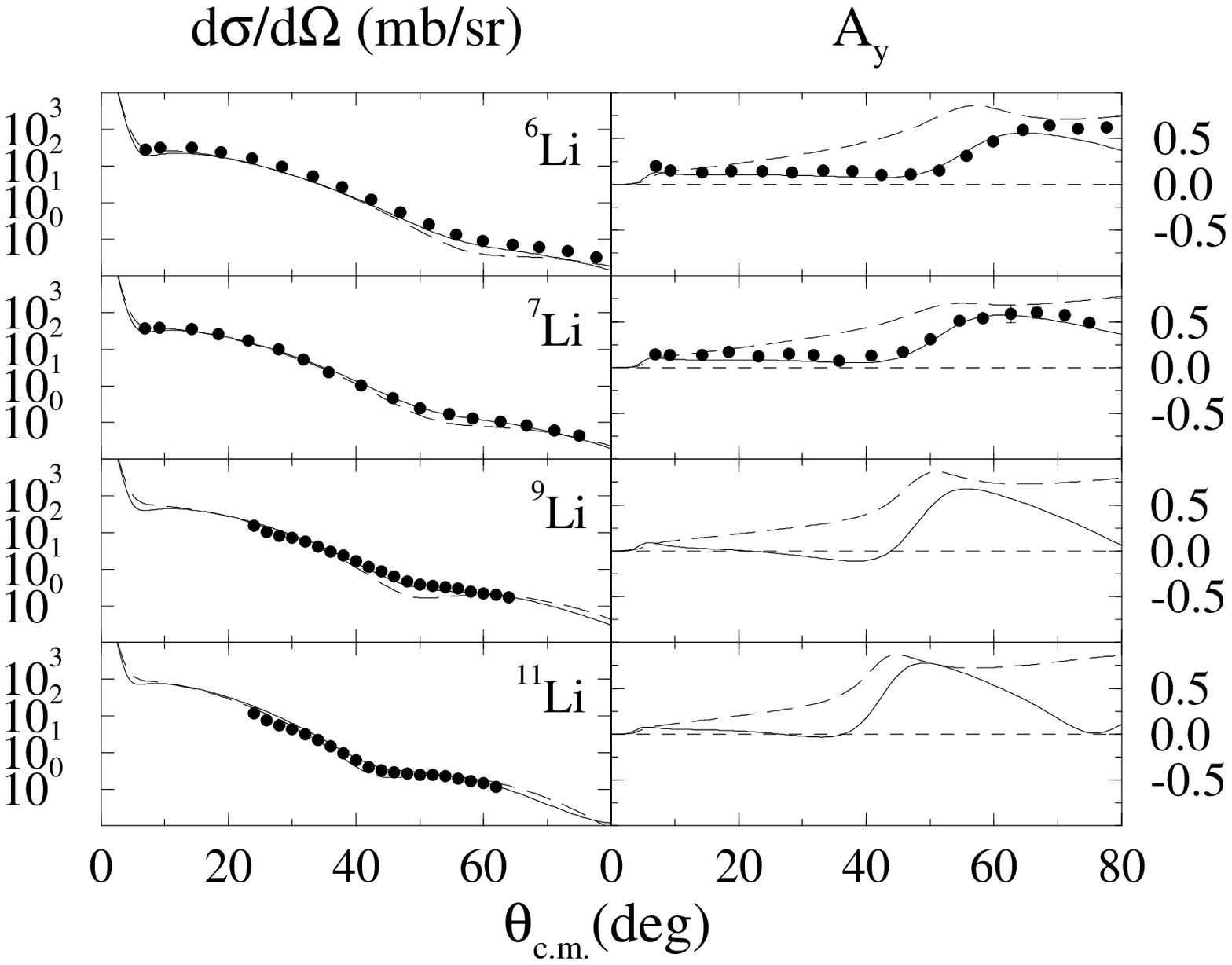,width=\linewidth}
\caption[*]{The differential cross sections (left) and analyzing
powers (right) from the elastic scattering of 65 MeV protons from
$^{6,7,9,11}$Li. The data (dots) are compared with the results of our
microscopic model calculations for the cases when medium effects are
included (solid curves) or are ignored (dashed curves).  Data were
taken from Refs.~\cite{Tosa85,Moon92}.}
\label{Li67911}
\end{figure}

\begin{figure}
\centering\epsfig{file=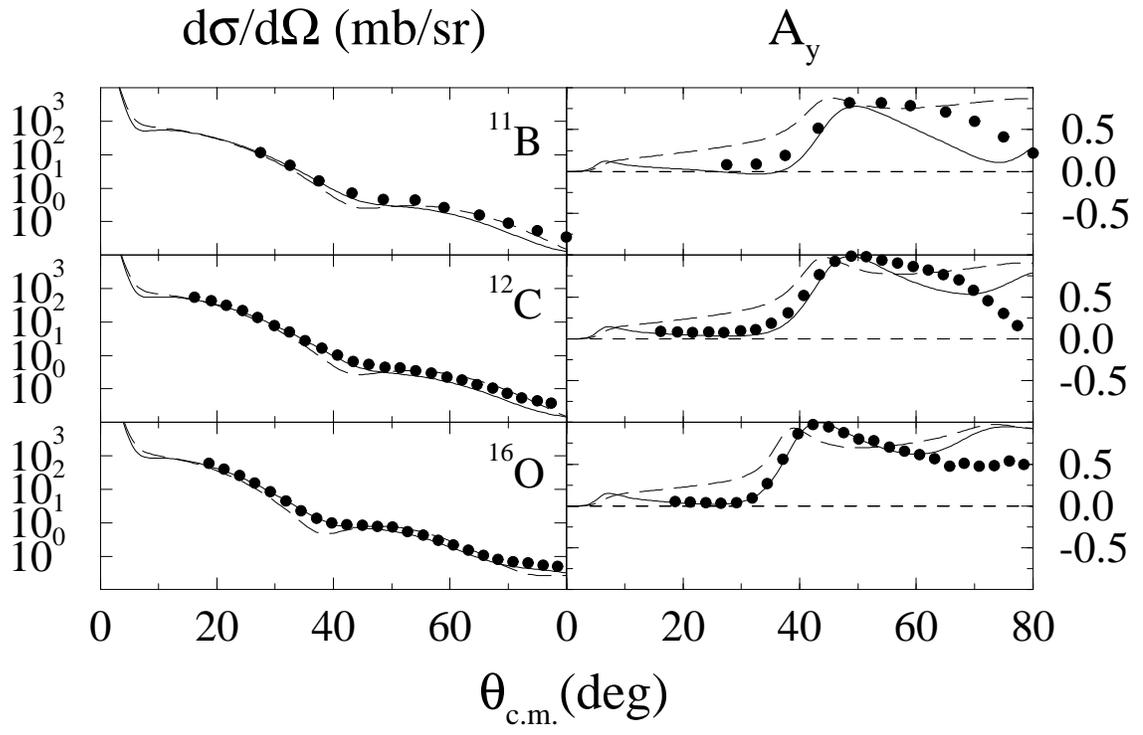,width=\linewidth}
\caption[*]{As for Fig.~\ref{Li67911} but for $^{11}$B, $^{12}$C and
$^{16}$O. Data were taken from
Refs.~\cite{Kuwa90,Yoso85,Kato85,Saka82}.}
\label{B11toO16}
\end{figure}

The results of our calculations of the elastic scattering of 65 MeV
protons from light mass nuclei are shown in Figs.~\ref{Li67911} and
\ref{B11toO16}. The differential cross sections, $d\sigma /d\Omega$,
and analyzing powers, $A_y$, are displayed in the left and right
panels, respectively. The target is identified in each segment of the
diagrams.  The ground state wave functions were obtained, for the most
part, from shell model calculations made within a complete
$(0+2)\hbar\omega$ model space.  The exceptions were $^{6,7}$Li, for
which the wave functions were obtained in a complete
$(0+2+4)\hbar\omega$ space~\cite{Li67}.  The SP wave functions were
assumed to be of Woods--Saxon (WS) form and were chosen to reproduce
the longitudinal elastic electron scattering form factor data. For
$^{16}$O, HO functions were used.

The differential cross sections and analyzing powers for the elastic
scattering of protons from four lithium isotopes are displayed in
Fig.~\ref{Li67911}. In the case of $^6$Li, our prediction of the cross
section made with the $g$--folding potential has the correct shape but
it is slightly weaker than the data. This is not the case at 200~MeV
\cite{Li67}, where the data are well reproduced. The slight
discrepancy at the lower energy may be a consequence of the
deformation of the target and the failure of the shell model, even in
the $(0+2+4)\hbar\omega$ space, to reproduce the cluster nature of the
$^6$Li nucleus~\cite{Li67}.  Yet, although $^6$Li does not have a
large central density, the changes as a result of the medium
modifications contained in the $g$ matrix are quite significant. This
is observed in comparison to the results obtained using the
$t$--folding potential.  The measured cross section is not reproduced
as well with that result as with the $g$--folding one, although the
discrepancies are not large. However, as noted
previously~\cite{C200,Arel95,Comm98}, the analyzing powers are very
sensitive to the details of the density in the folding. We find this
to be the case again, and at 65 MeV, with all of the light nuclei. Our
results of scattering from $^7$Li are better, as now our prediction
for the cross section obtained using the $g$--folding potential are
only marginally weaker than the data.  The analyzing power is also
well reproduced out to 80$^\circ$.

For $^9$Li, the medium modification effects on the cross section are
quite pronounced. In this case, it is essential to use the results
from $g$--folding. Currently there are no data for the spin
observables or for form factors from electron scattering for the
exotic nuclei so that the proton cross section represents the best
available test of putative structures. In this case, as for $^{11}$Li,
the density profile was obtained directly from the shell model ground
state. Also, in absence of any electron scattering data, we used WS SP
wave functions appropriate for electron scattering from $^9$Be. With
such a specification, the result obtained from the $g$--folding
potential is remarkably good.

The final results shown in Fig.~\ref{Li67911} are for $^{11}$Li as the
target. This nucleus has a halo distribution associated with the very
loosely bound valence neutron pair. To allow for this extension in the
density we use WS SP wave functions with the binding energy for the
$0p_{\frac{1}{2}}$ and higher shell neutrons set to 500 keV
\cite{Li911ref}. The WS functions assumed for the $0s_{\frac{1}{2}}$
and $0p_{\frac{3}{2}}$ orbits were those used for $^9$Li, consistent
with the model for $^{11}$Li of a two-neutron halo outside a $^9$Li
core. With that prescription, our predictions adequately reproduce the
data, although it slightly over-estimates the cross section at the
small scattering angles. Even so, using such a simple model to
describe the halo nature of this nucleus produces results that give
confidence to use proton scattering to study the matter distributions
of other neutron and proton rich nuclei.

We note in passing that for a target with non--zero spin it is
essential to include all multipole contributions in the scattering, as
is the case with all the lithium isotopes. In Eq.~(\ref{obdme}), there
are $2J+1$ possible multipoles in the specification of the ground
state.  The contributions of non-zero multipoles are not large, but
they are important. The analyzing power is especially sensitive to
them, as is most dramatically seen in the results for proton
scattering from $^9$Be~\cite{mass200,Be9ref}.

The results for scattering from $^{11}$B, $^{12}$C and $^{16}$O are
presented in Fig.~\ref{B11toO16}. For $^{11}$B, our predictions
compare least favourably with the data, although the cross section and
analyzing power still are well described out to $40^\circ$. Above
this, the predicted cross section falls at a slightly greater rate
than that indicated by the data. However, the effects of the medium
modifications in the effective $NN$ interaction are essential for any
adequate description of the analyzing power. With the other nuclei,
the $g$--folding model predictions are all very good. They are clearly
better representations of the data, especially the analyzing powers,
than are the results found with the $t$--folding model. Note that the
range involves at least three orders of magnitude in the cross
section.

\subsection{Medium mass nuclei (16 $<$ A $\leq$ 64)}

While for light mass nuclei it is now possible to make shell model
calculations using large and complete spaces, for $1s 0d$ and $0f 1p$
shell nuclei, such ``no--core'' determination of the shell model wave
functions as yet are not feasible.  The size of the spaces necessary
are prohibitively large. For a number of nuclei, however, it is still
possible to construct complete $0\hbar\omega$ shell models for use in
analyses of elastic scattering data. Those models still require core
polarization corrections and and hence the use of the density matrices
in analyses of electron scattering data may not be as good a check on
the SP wave functions as with the $0p$--shell nuclei.  Therefore we
have assumed HO functions with $A^{1/6}$~fm for all subsequent
calculations of the proton scattering observables for nuclei with $A
\geq 20$.

\begin{figure}
\centering\epsfig{file=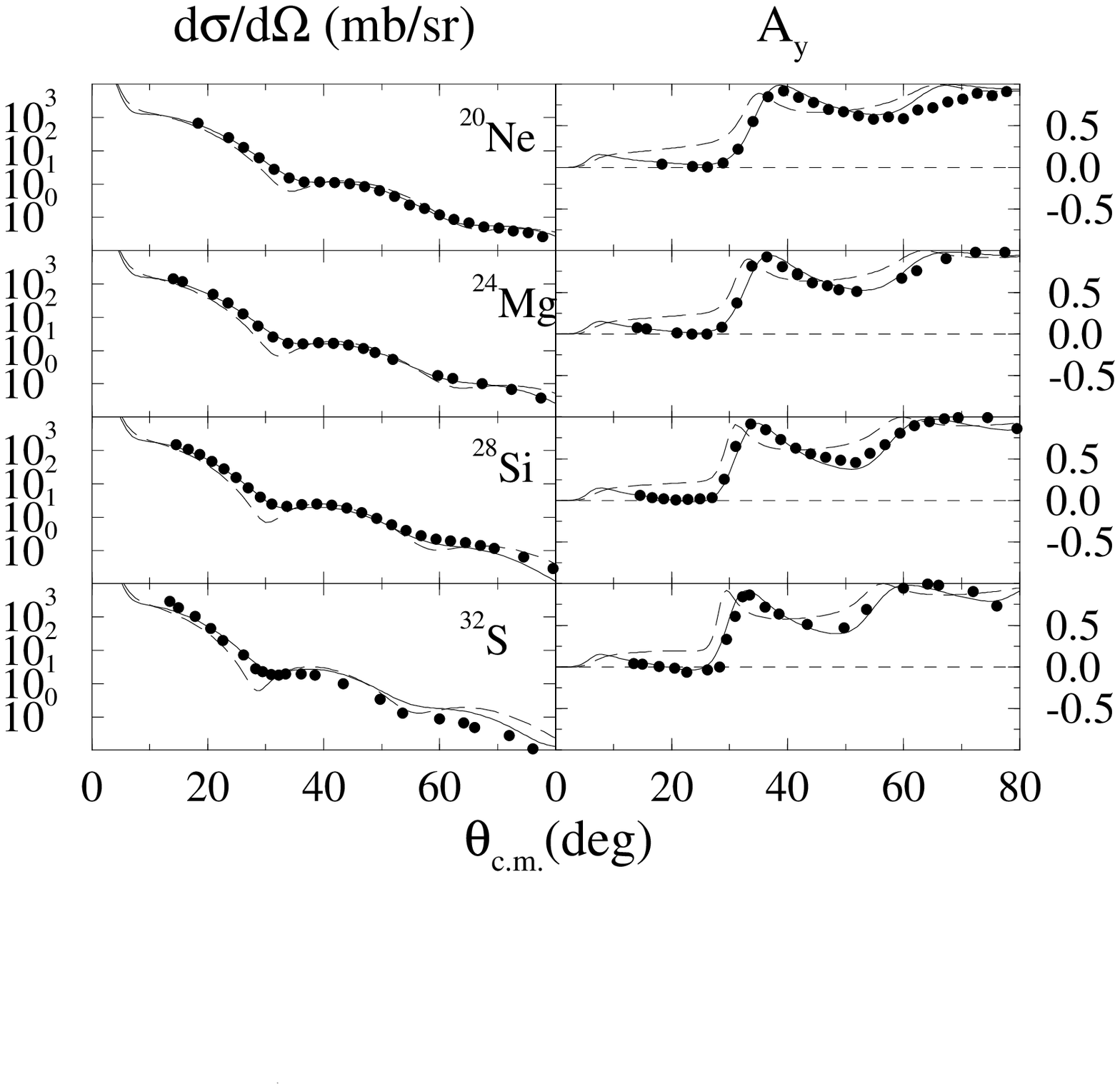,width=\linewidth}
\caption[*]{As for Fig.~\ref{Li67911} but for $^{20}$Ne, $^{24}$Mg,
$^{28}$Si and $^{32}$S. Data were taken from
Refs.~\cite{Kato85,Saka82}.}
\label{Ne20toS32fig}
\end{figure}

The results for the elastic scattering of 65 MeV protons from
$^{20}$Ne, $^{24}$Mg, $^{28}$Si and $^{32}$S are presented in
Fig.~\ref{Ne20toS32fig}. A complete $0\hbar\omega$ ($1s0d$) shell
model calculation was performed for all nuclei.  The results of our
$g$--folding model calculations reproduce the cross--section data for
the scattering from $^{24}$Mg and $^{28}$Si well and at all scattering
angles to $80^\circ$.  For $^{20}$Ne, the description is accurate to
$60^\circ$ at least. However for scattering from $^{32}$S, the level
of agreement with the cross--section data is less than
satisfactory. Also in all cases, the level of agreement between the
results of our calculations using the $t$--folding model and the data
is poorer.  This poor reproduction we find from our $g$--folding model
to the $^{32}$S cross--section data cannot be due simply to the choice
of oscillator length. The cross--section data for $^{32}$S have
magnitudes considerably larger than those of $^{28}$Si (and of
$^{40}$Ar that we show next) for small scattering angles and
vice--versa for larger scattering angles.  It is known that these $1s
0d$ shell nuclei are deformed and that ${}^{32}$S is different to the
others.  That is evidenced by ${}^{32}$S not having the distinctive
splitting of the giant dipole resonances that occur in ${}^{24}$Mg and
${}^{28}$Si.  The different deformation of ${}^{32}$S might explain
the difference we see in the quality of reproduction of the scattering
data.  Certainly when a phenomenological optical model analysis was
used to fit the same scattering data, the parameters required to fit
the ${}^{32}$S data were quite different from those found with data
off the neighbouring nuclei~\cite{Kato85}.  As with the results
presented for the light nuclei, the differences between the results
obtained using the $g$ and $t$--folding models are far more
significant in the analyzing powers. In all cases, those results
obtained with the $g$--folding models reproduce the data well. We also
note that there is a definite trend in the size and shape of the data
as one increases the mass of the target. The data indicate a sharp
rise from 0 at $0^{\circ}$ to a maximum near 1 followed by a fall and
a second peak of similar character. The first peak becomes more
forward-peaked as the mass of the target increases ($60^{\circ}$ for
$^6$Li, $30^{\circ}$ for $^{28}$Si). This could be used as an
indicator for optimum SP wave functions, as small changes in the HO
parameter produce shifts in the position of this peak in the analyzing
power.

\begin{figure}
\centering\epsfig{file=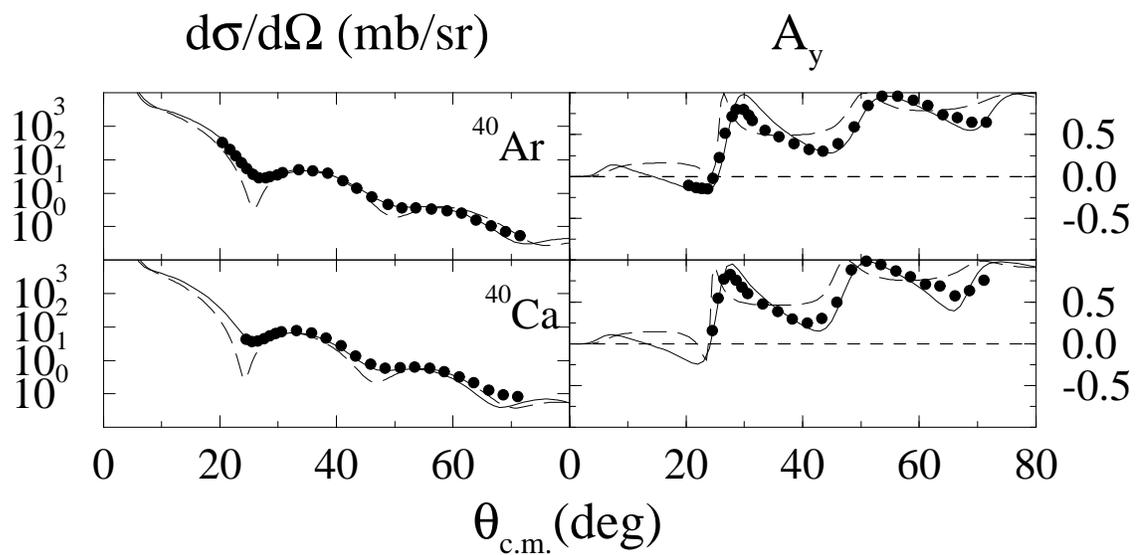,width=\linewidth}
\caption[*]{As for Fig.~\ref{Li67911} but for $^{40}$Ar and
$^{40}$Ca. Data were taken from Refs.~\cite{Kato85,Saka82,Noro81}.}
\label{Ar40toCa40}
\end{figure}

The results for the elastic scattering of 65 MeV protons from
$^{40}$Ar and $^{40}$Ca are displayed in Fig.~\ref{Ar40toCa40}.  The
analyzing powers have small differences with the most notable being a
small shift in the angles of structures.  The differential cross
sections also are different with the position and size of the
prominent peak being the primary effect.  For ${}^{40}$Ar that feature
occurs at 33$^\circ$ and has a value of 60 mb/sr.  For ${}^{40}$Ca,
the result is 44 mb/sr at 36$^\circ$.  For both these nuclei, a
`packed' model of their structure has been used in our analyses.
Specifically ${}^{40}$Ca has been taken as a doubly closed shell
nucleus while ${}^{40}$Ar has been taken as two proton holes (in the
$d_{\frac{3}{2}}$-subshell) with two extra core nucleons (in the
$f_{\frac{7}{2}}$ orbit) on that doubly closed shell description.  The
agreement with data for the scattering from $^{40}$Ar is good to
$60^\circ$.  The differences between the Ar and Ca results are
consistent with observed differences in the data sets and reflect the
effects of surface contributions revealed by a change of basic
structure from the closure of a major shell.

In the case of $^{40}$Ca, our $t$--folding model results are quite
similar to those obtained recently in a momentum-space
framework~\cite{Elst98}, while our $g$--folding model results agree
quantitatively with those found using a $g$ matrix in another momentum
space calculation~\cite{Arel95}.  These differences emphasise that one
cannot neglect the importance of the medium in specifying the
effective interaction, whether it is for scattering at 200
MeV~\cite{C200} or at 65 MeV~\cite{Comm98}.

\begin{figure}
\centering\epsfig{file=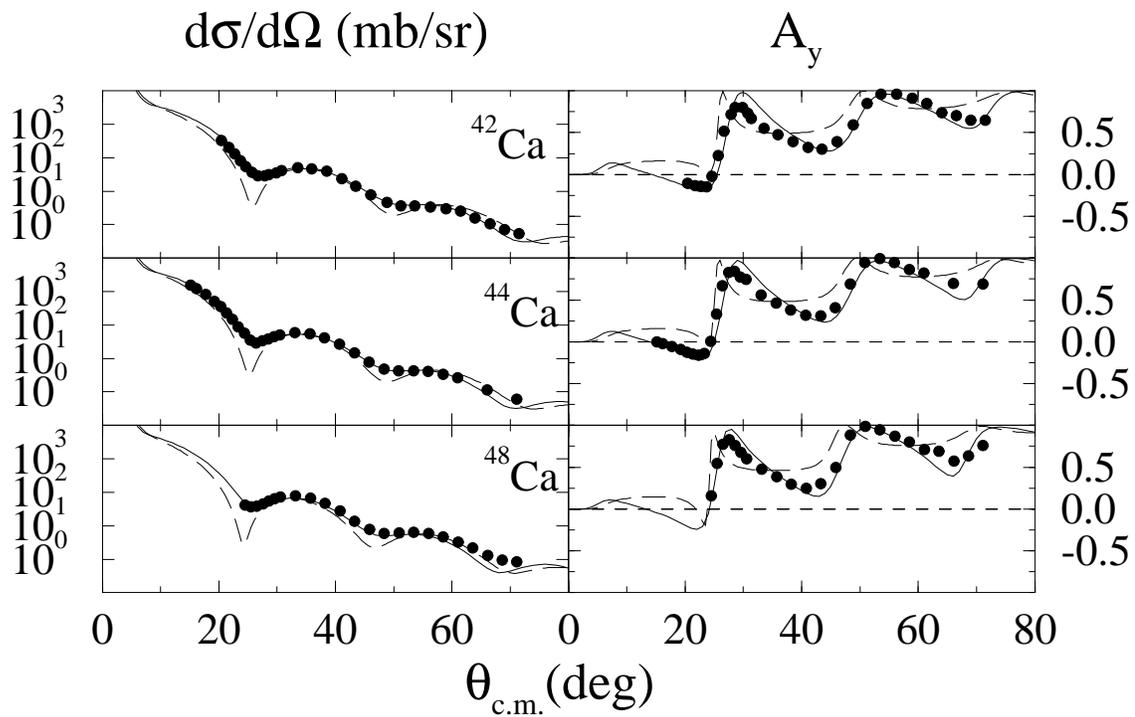,width=\linewidth}
\caption[*]{As for Fig.~\ref{Li67911} but for $^{42,44,48}$Ca.  Data
were taken from Refs.~\cite{Saka82,Noro81}.}
\label{Ca424448fig}
\end{figure}

The results of our calculations using the $g$-- and $t$--folding
models are compared with data for the case of scattering from the
heavier calcium isotopes in Fig.~\ref{Ca424448fig}. As with the
lighter nuclei, we again note that the results found using the
optical potentials specified in the $g$--folding model are in
excellent agreement with the data up to $70^{\circ}$. Again those
results differ markedly from ones obtained using the optical
potentials in the $t$--folding model. Also, the trend in the
analyzing power noted earlier is observed in the calcium
isotopes. However, there are more pronounced minima in the cross
sections, and the analyzing powers show a new minimum developing at
small scattering angles.

\begin{figure}
\centering\epsfig{file=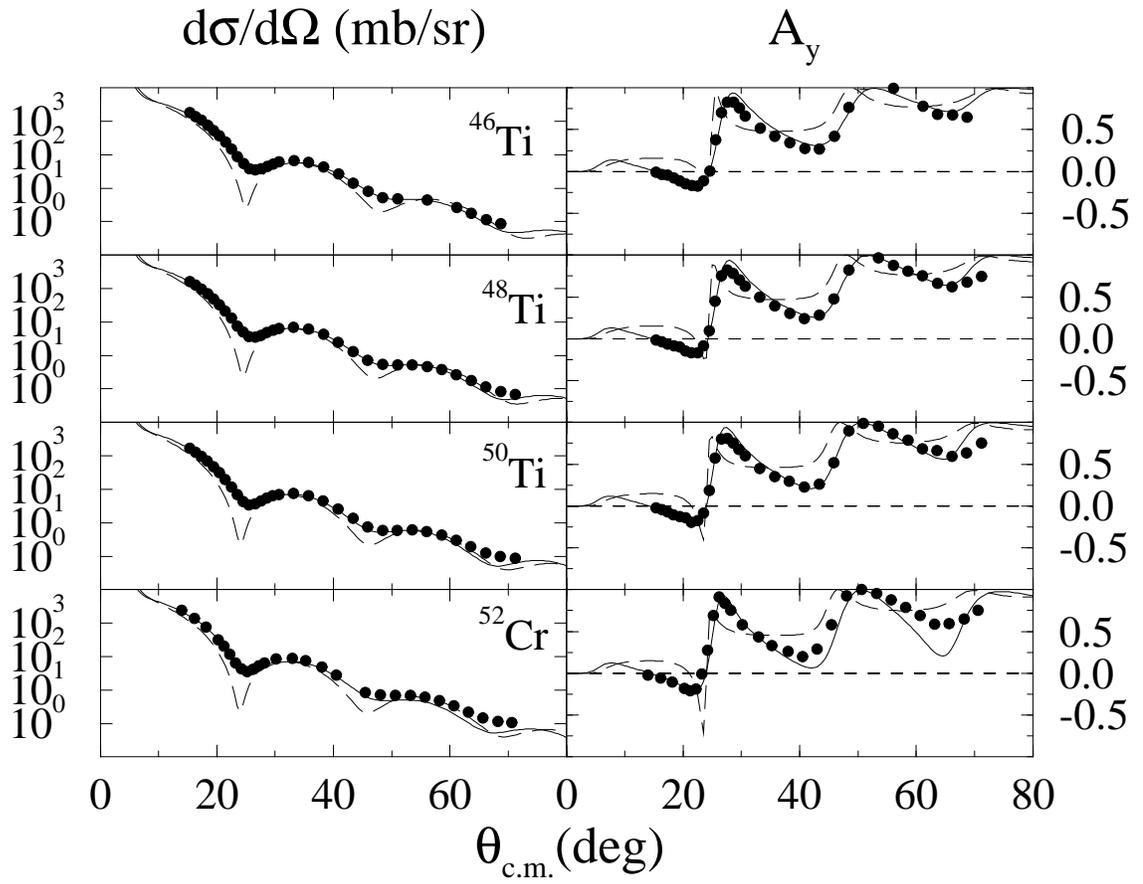,width=\linewidth}
\caption[*]{As for Fig.~\ref{Li67911} but for $^{46,48,50}$Ti and
$^{52}$Cr. Data were taken from Refs.~\cite{Saka82,Noro81}.}
\label{TitoCr52fig}
\end{figure}

\begin{figure}
\centering\epsfig{file=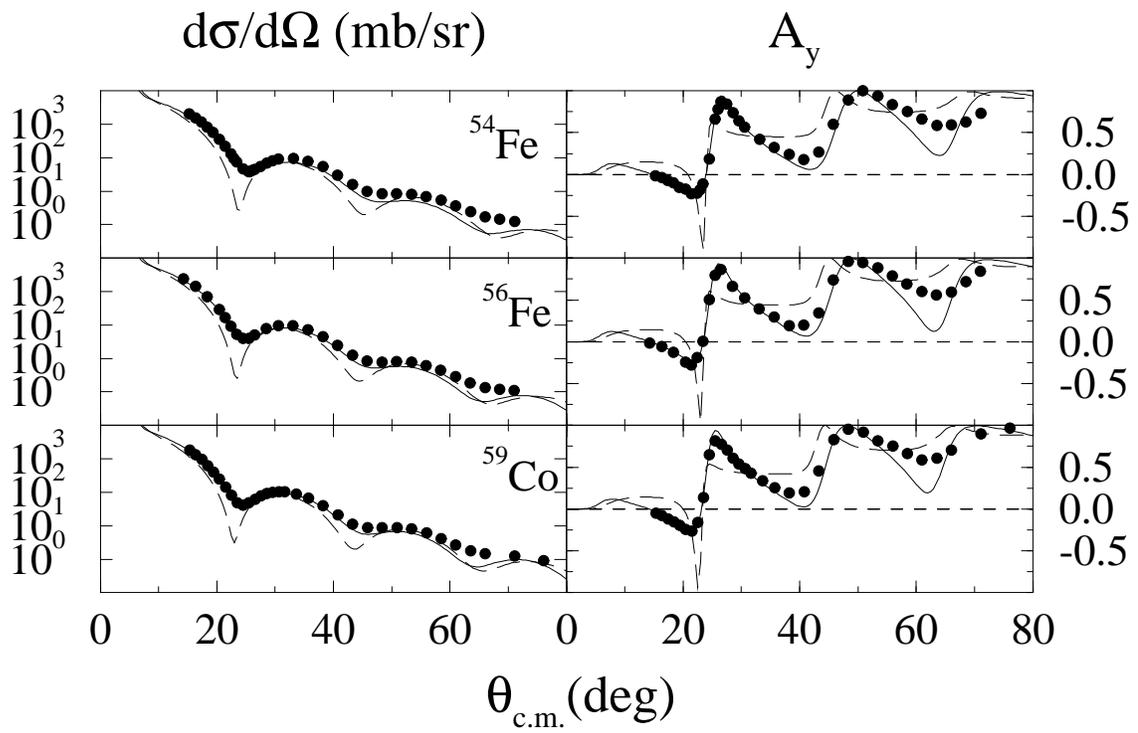,width=\linewidth}
\caption[*]{As for Fig.~\ref{Li67911} but for $^{54,56}$Fe and $^{59}$Co.
Data were taken from Ref.~\cite{Saka82}.}
\label{Fe54toCu63fig}
\end{figure}

\begin{figure}
\centering\epsfig{file=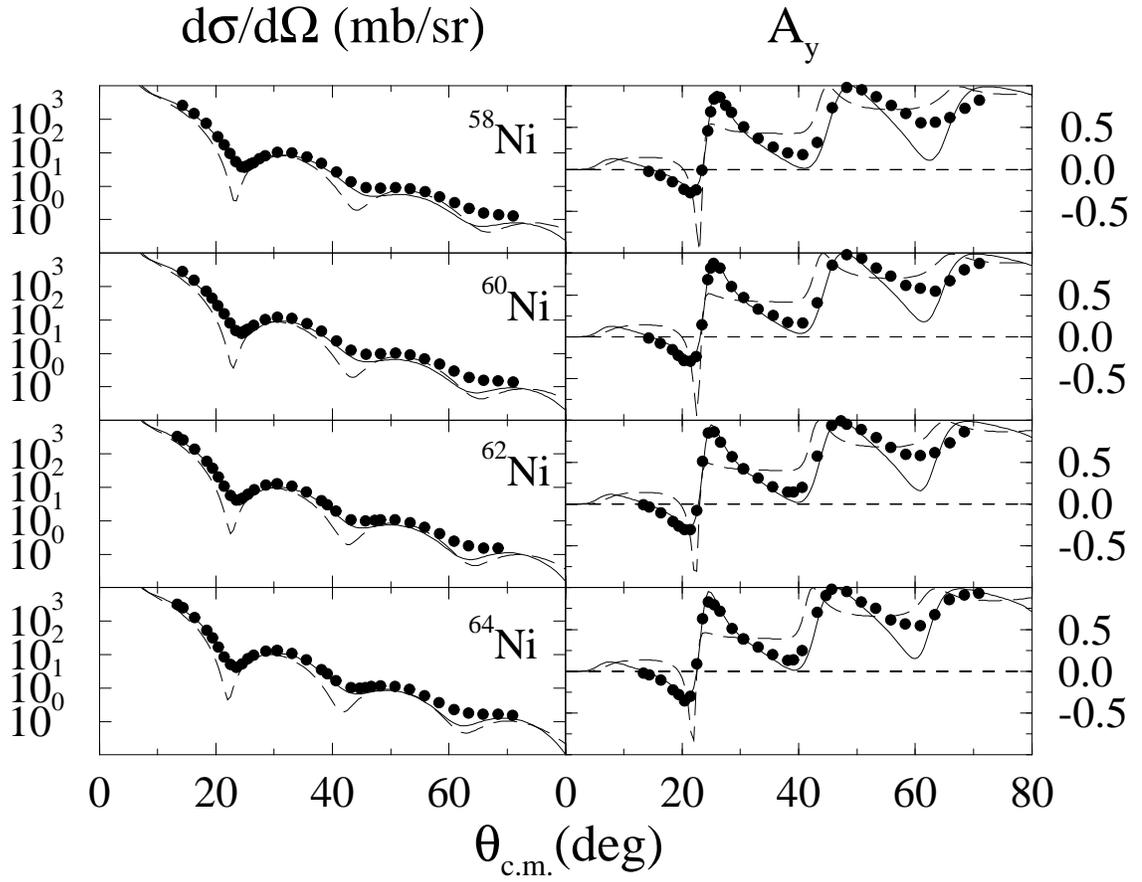,width=\linewidth}
\caption[*]{As for Fig.~\ref{Li67911} but for $^{58,60,62,64}$Ni.
Data were taken from Ref.~\cite{Saka82}.}
\label{Ni58606264fig}
\end{figure}

The results for the scattering from $^{46,48,50}$Ti and $^{52}$Cr are
displayed in Fig.~\ref{TitoCr52fig}. The agreement between the data
for the scattering of 65 MeV protons from the titanium isotopes and
the results from our $g$--folding model calculations is quite good to
$70^{\circ}$. That level of agreement is not observed in the case of
scattering from $^{52}$Cr; the data are under--predicted above
$45^{\circ}$. However, the shape indicated by the data is by far
better reproduced by the $g$--folding result than the $t$--folding
one. This under--prediction at large angles is observed in the
scattering from $^{54,56}$Fe and $^{59}$Co, displayed in
Fig.~\ref{Fe54toCu63fig}, and also in the scattering from the Ni
isotopes, displayed in Fig.~\ref{Ni58606264fig}. In these cases, a
complete $0\hbar\omega$ shell model calculation is not possible with
the standard diagonalization techniques; those are only calculable in
an approximate fashion using Monte Carlo techniques. For the present
calculations, the model space in the shell model calculation was
restricted to close the $0f_{\frac{7}{2}}$ orbit. This could cause the
observed discrepancy with the data. Also, the choice for the
oscillator length might not be optimal.  One could choose to set $b$
by a best fit to the cross--section and analyzing power data, as we
did in our 200 MeV analysis~\cite{mass200}.  Nevertheless, in every
case, the results obtained in the $g$--folding model are in much
better agreement overall with the data, especially with the analyzing
powers, with the level of agreement being actually quite good.

The trends in magnitudes and shape of the data with mass remain and
are enhanced with this set of nuclei.  Notably the first minimum of
the cross section becomes more pronounced as does the first minimum
in the analyzing power.

\subsection{Heavy nuclei (A $>$ 64)}

For nuclei with mass in excess of 70, shell model calculations, even
in restricted spaces, generally are not feasible. The increase in the
number of valence nucleons, and an increase in the number of SP states
that may be occupied, cause the dimension of the model space to become
prohibitively large.  Therefore, we have chosen a simple packing model
to specify the OBDME.  In that model, nuclear shells are filled in
sequence from the lowest lying shells to the Fermi level.  As the
results of our calculations are not i particularly sensitive to the
diffuseness in the nuclear density in this mass region, this is not a
bad approximation.  A more important feature is the choice of the
oscillator length for the single nucleon bound state functions.
Again, we have chosen $b = A^{1/6}$ fm for all shells. A more
reasonable approach would be to assume a different oscillator
parameter for the protons and neutrons.  By that means, the proton and
neutron total distributions could be kept similar despite the neutron
excesses.

\begin{figure}
\centering\epsfig{file=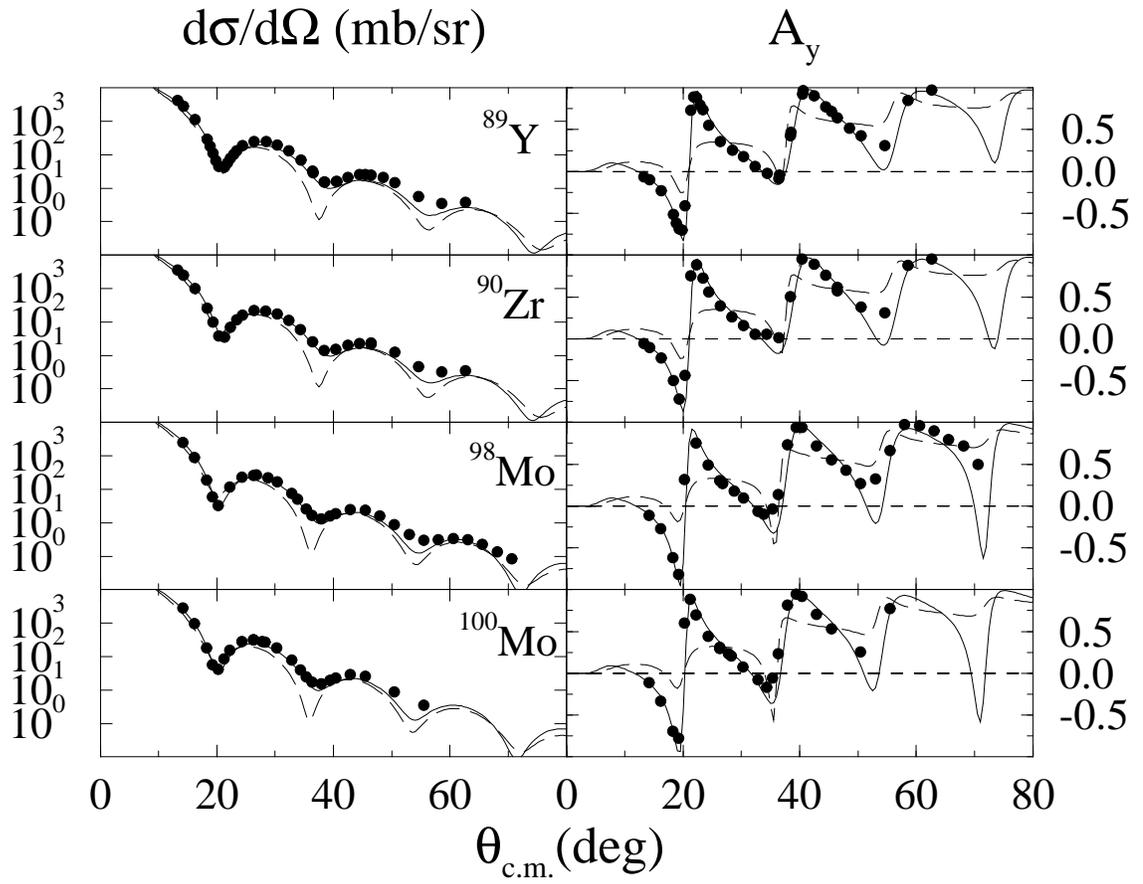,width=\linewidth}
\caption[*]{As for Fig.~\ref{Li67911} but for $^{89}$Y, $^{90}$Zr
and $^{98,100}$Mo.
Data were taken from Ref.~\cite{Saka82}.}
\label{Y89toMo100fig}
\end{figure}

The results for the scattering from $^{89}$Y, $^{90}$Zr and
$^{98,100}$Mo are displayed in Fig.~\ref{Y89toMo100fig}. For the four
cases presented, the cross-section data are well reproduced by our
$g$--folding model predictions to $50^{\circ}$. Thereafter, our
results slightly underestimate the data in the region of the minimum
at $\sim 55^{\circ}$. While the $t$--folding model results give
similar shapes for the cross sections, the second minimum at
$35^{\circ}$ is an order of magnitude deeper than that observed and
also as predicted by the $g$--folding model result. The differences
between the two models are far more striking in the comparisons of the
analyzing powers. The $g$--folding results are in very good agreement
with the data to $50^{\circ}$. The results obtained with the
$t$--folding model definitely are not. The latter do not reproduce the
shape or the magnitude of the data. It is interesting to note that the
region in which the analyzing power is underestimated by the
$g$--folding results is also that in which the cross section is
underestimated. Since the analyzing power is scaled by the cross
section, an improvement in the level of agreement in the cross section
in this region may also produce an improvement in the analyzing power.

\begin{figure}
\centering\epsfig{file=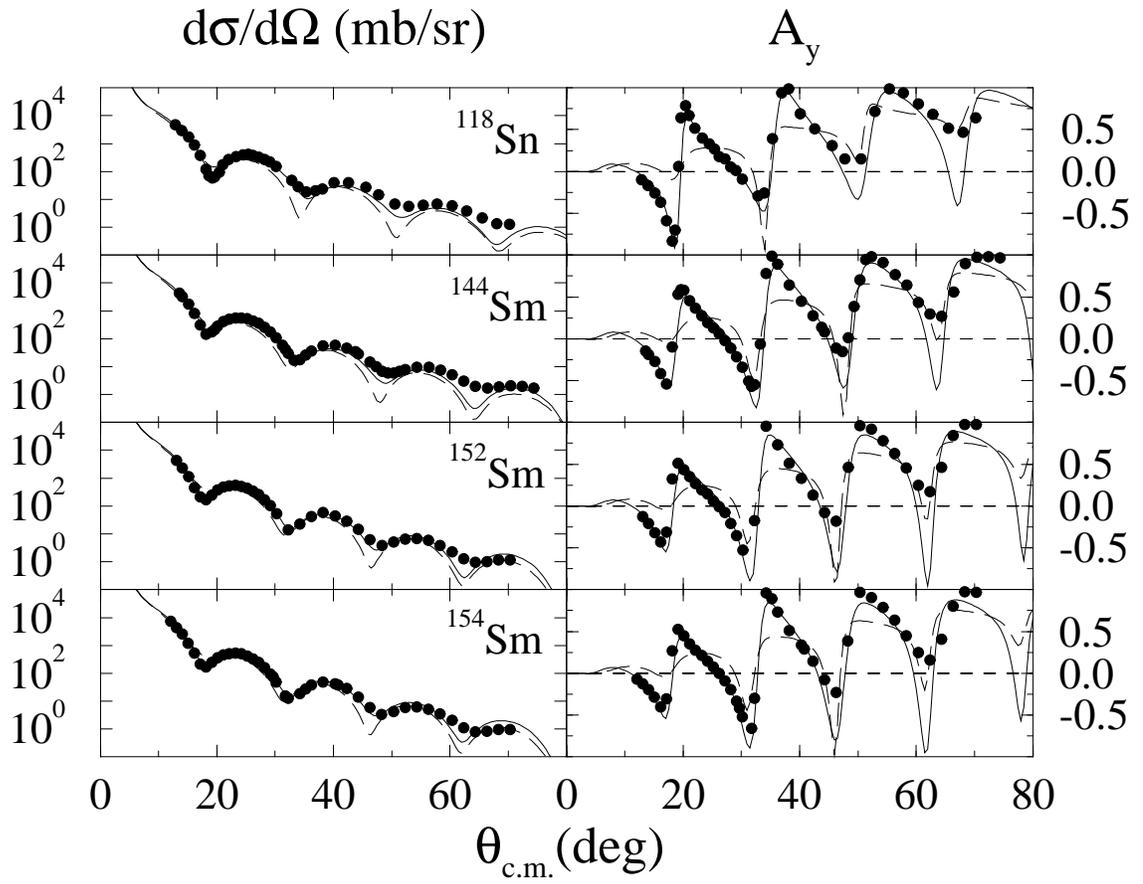,width=\linewidth}
\caption[*]{As for Fig.~\ref{Li67911} but for $^{118}$Sn and
$^{144,152,154}$Sm.  Data were taken from
Refs.~\cite{Yoso85,Saka82,Ohta83}.}
\label{Sn118toSm154fig}
\end{figure}

Our $g$--folding model results are compared with data for the
scattering of 65~MeV protons from $^{118}$Sn and $^{144,152,154}$Sm in
Fig.~\ref{Sn118toSm154fig}. For these cases, the first two minima in
the cross-section data are very well reproduced by the $g$--folding
model results as is the third minimum in the $^{152,154}$Sm data. The
same level of agreement is achieved in the analyzing powers between
the data and results from the $g$--folding calculations. The same
dramatic difference between the $g$-- and $t$-- models is observed in
the analyzing power as was the case with the results in the mass-90
region.

\begin{figure}
\centering\epsfig{file=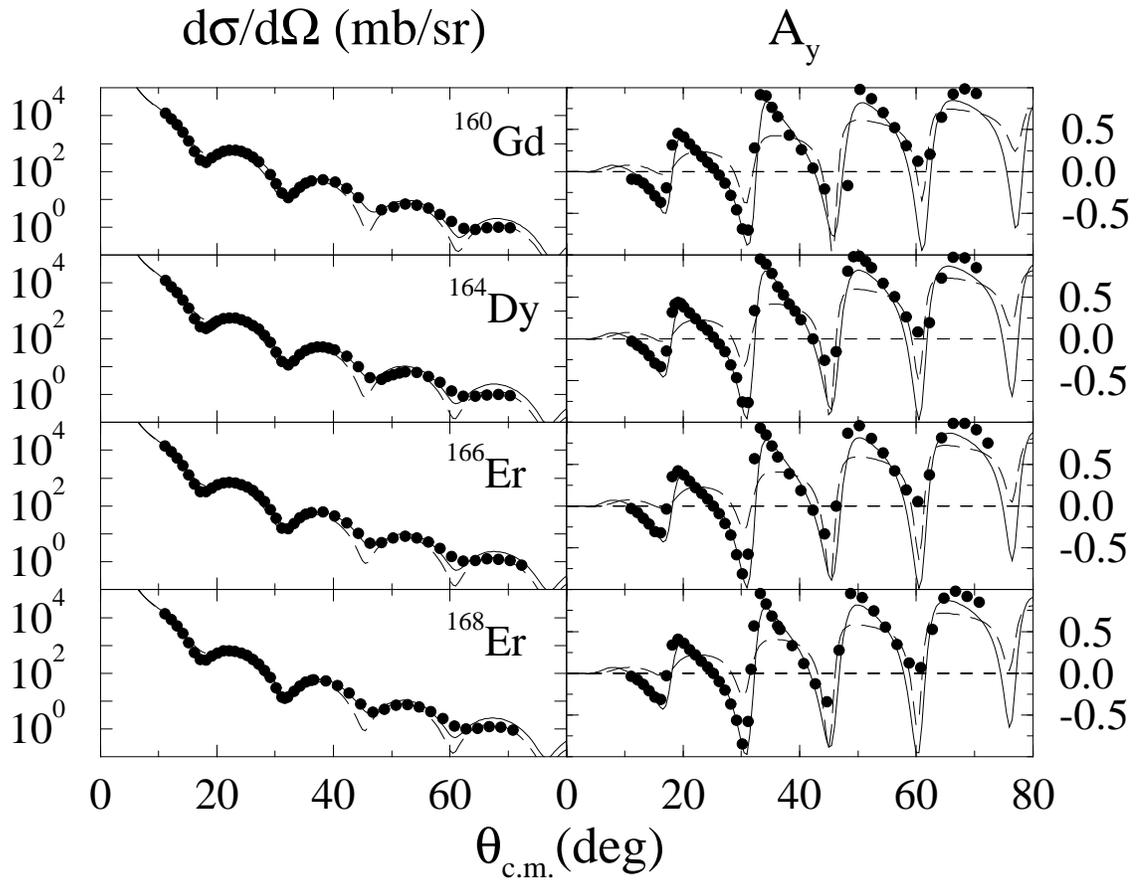,width=\linewidth}
\caption[*]{As for Fig.~\ref{Li67911} but for $^{160}$Gd, $^{164}$Dy
and $^{166,168}$Er.  Data were taken from
Refs.~\cite{Ohta83,Ichi87,Ichi84}.}
\label{Gd160toEr168fig}
\end{figure}

\begin{figure}
\centering\epsfig{file=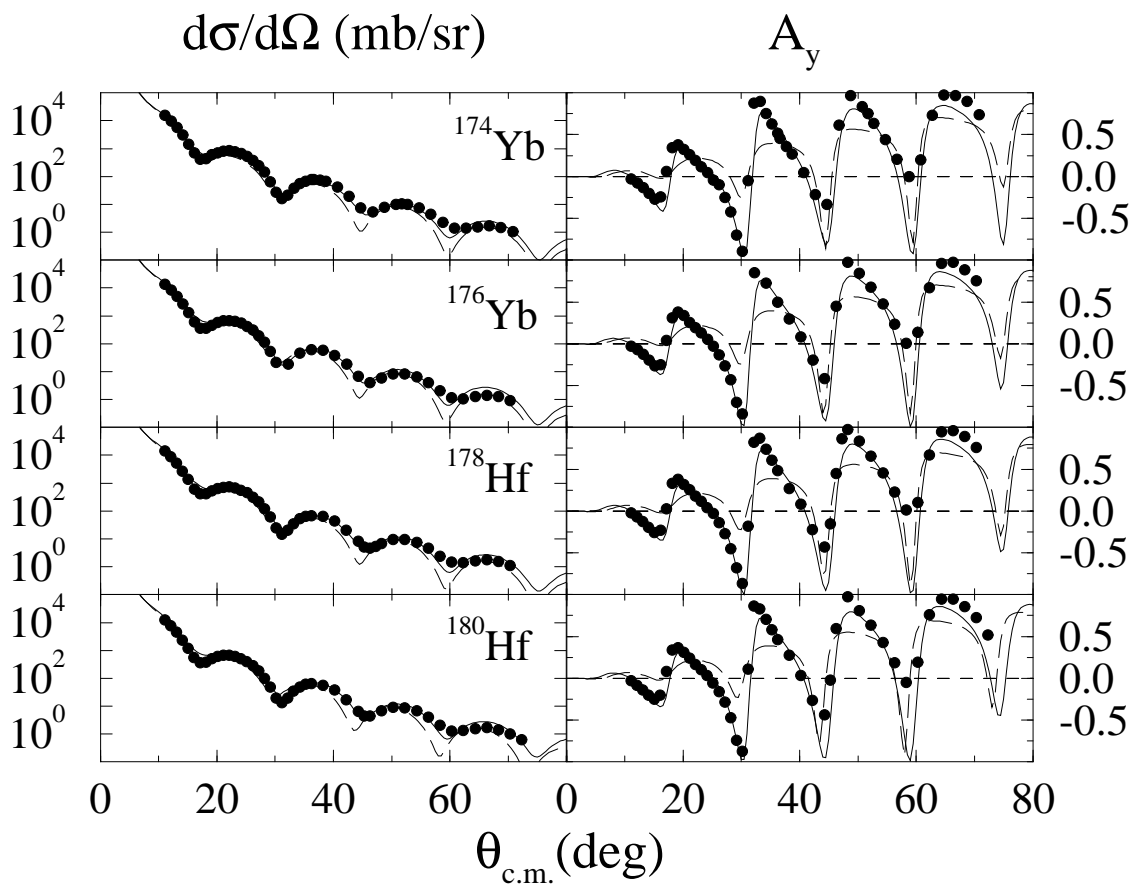,width=\linewidth}
\caption[*]{As for Fig.~\ref{Li67911} but for $^{174,176}$Yb
and $^{178,180}$Hf.
Data were taken from Refs.~\cite{Ichi84,Ogaw86}.}
\label{Yb174toHf180fig}
\end{figure}

We show comparisons of our model predictions with data for the
scattering from nuclei ranging between $^{160}$Gd and $^{180}$Hf in
Figs.~\ref{Gd160toEr168fig} and \ref{Yb174toHf180fig}. The level of
agreement in the cross section between the data and the $g$--folding
model results is again very high. The results obtained using the
$t$--folding model have a tendency to underestimate the data,
especially at the minima above $40^{\circ}$, and do not predict the
locations of those minima.  Nevertheless the shape of the
cross--section data generally is reproduced.  Those levels of
agreement are not reflected in the comparisons of analyzing
powers. The $g$--folding model predictions are results that generally
reflect the data, although there might be some suggestion from the
comparisons with the low angle data that larger oscillator lengths are
preferable. The $t$--folding model ones reproduce neither the
magnitudes nor the shapes of the data.

\begin{figure}
\centering\epsfig{file=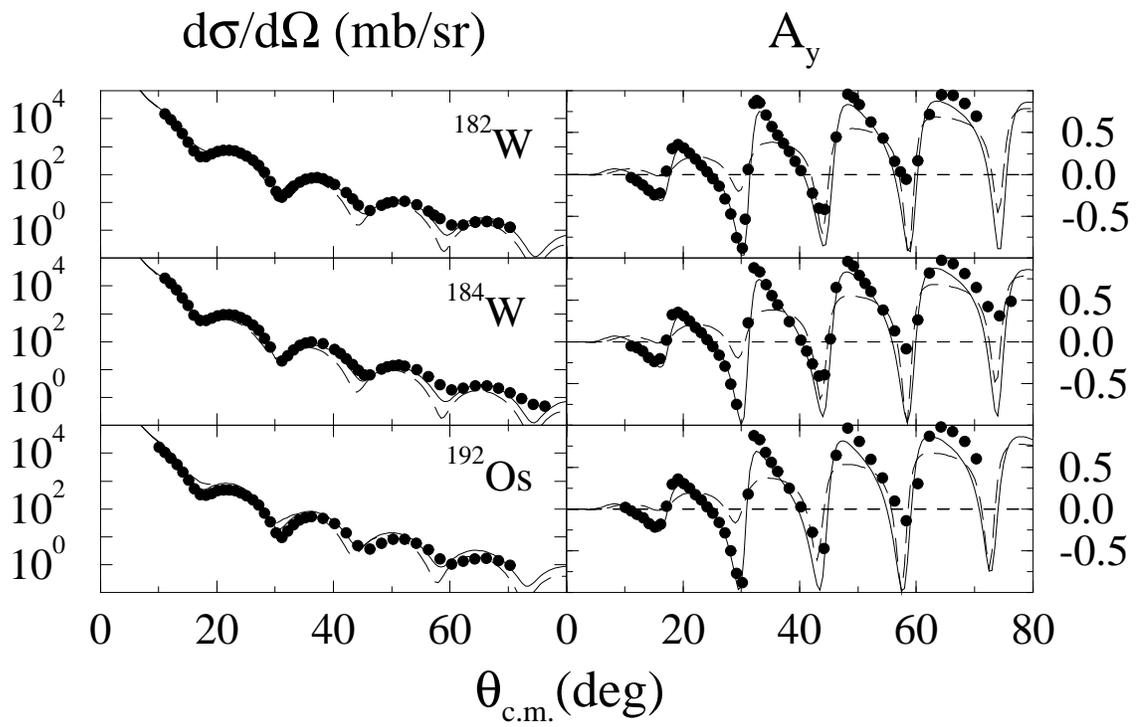,width=\linewidth}
\caption[*]{As for Fig.~\ref{Li67911} but for $^{182,184}$W
and $^{192}$Os. Data were taken from Refs.~\cite{Ichi87,Ogaw86}.}
\label{W182toAu197fig}
\end{figure}

The data for the scattering from $^{182,184}$W, and $^{192}$Os are
compared with our $g$-- and $t$--folding model predictions in
Fig.~\ref{W182toAu197fig}. The level of agreement with the
cross--section and analyzing power data for the scattering from the W
isotopes is as observed for the cases discussed already.  The results
for the scattering from $^{192}$Os are a little perplexing.  The data
suggest a somewhat weaker cross section than those for the elastic
scattering from neighbouring nuclei.  Yet the shape and magnitude of
the analyzing power is similar. Our $g$--folding results overestimate
the cross section by $60-70$\% at low scattering angles.

\begin{figure}
\centering\epsfig{file=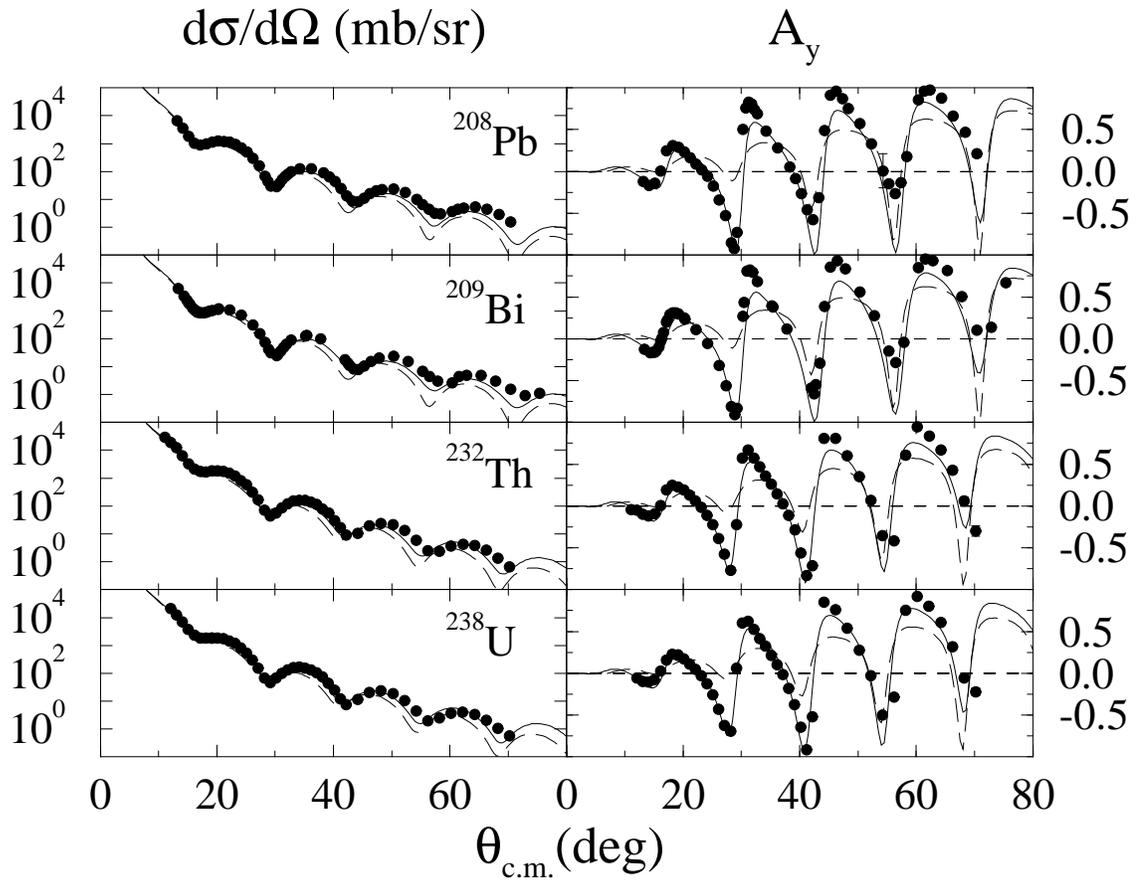,width=\linewidth}
\caption[*]{As for Fig.~\ref{Li67911} but for $^{208}$Pb, $^{209}$Bi,
$^{232}$Th and $^{238}$U.  Data were taken from
Refs.~\cite{Saka82,Take86}.}
\label{Mass200fig}
\end{figure}

We compare the results of our microscopic optical model calculations
with the data for scattering from nuclei with $A > 200$ in
Fig.~\ref{Mass200fig}. All the data, for both the cross section and
analyzing power, show similar structure, with which our $g$-folding
model results agree well in both shape and magnitude.  However, the
slight differences in the positions of the minima between the
cross--section data and our predictions suggest that the choice of
oscillator wave functions is less than optimal.  Yet the comparison
with the analyzing power data is remarkably good in tracking the shape
and positions of the minima.  It is clear once more that the effects
of the nuclear medium in defining the effective $NA$ optical potential
are quite important. While the $t$-folding model results track the
positions of the maxima and minima in the analyzing powers to some
extent, they fail to reproduce the observed magnitudes. This is also
reflected in the cross sections, where the positions of the minima
also are not reproduced.

\subsection{Mass dependencies of spin observables}

\begin{figure}
\centering\epsfig{file=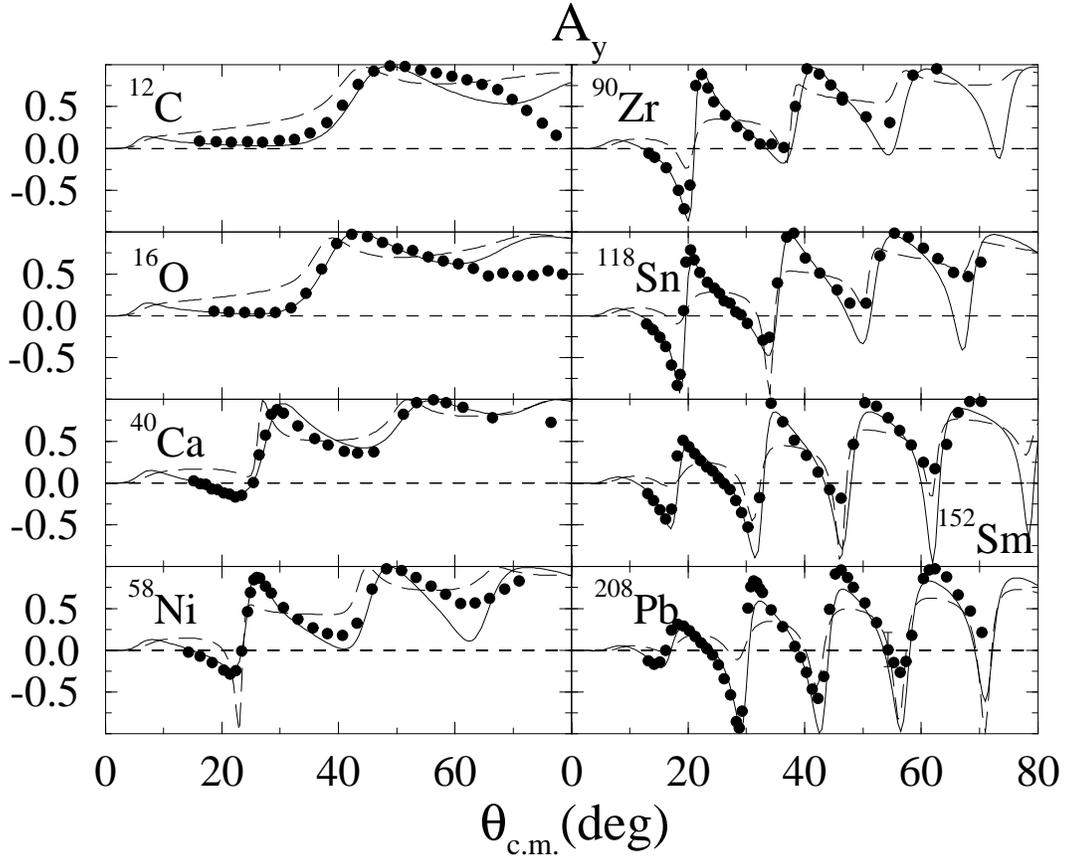,width=\linewidth}
\caption[*]{The analyzing power, A$_y$, from the elastic scattering of
65 MeV protons from $^{12}$C, $^{16}$O, $^{40}$Ca, $^{58}$Ni, $^{90}$Zr,
$^{118}$Sn, $^{154}$Sm and $^{208}$Pb. The data (dots) are
compared with the results of our microscopic model calculations
for the cases when medium effects are included (solid curves) or
are ignored (dashed curves).
Data were taken from Ref.~\cite{Yoso85,Kato85,Saka82,Ohta83}.}
\label{SpinAfig}
\end{figure}

\begin{figure}
\centering\epsfig{file=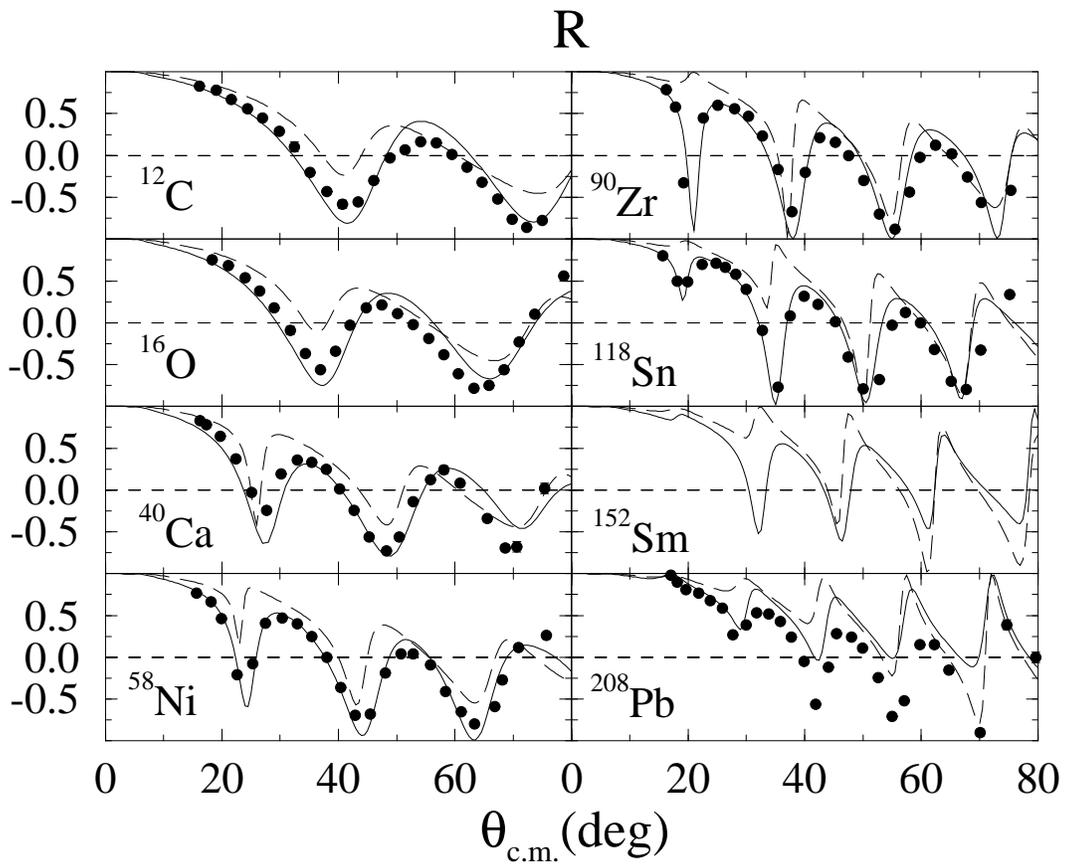,width=\linewidth}
\caption[*]{As for Fig.~\ref{SpinAfig} but for the spin rotation, R.
Data were taken from Ref.~\cite{Yoso85}.}
\label{SpinRfig}
\end{figure}

In Figs.~\ref{SpinAfig} and \ref{SpinRfig} we display the analyzing
powers and spin rotations ($R$) for the scattering of 65~MeV polarized
protons from a set of 8 nuclei ranging from $^{12}$C to
$^{208}$Pb. The curves represent the same model predictions as given
in the preceding figures. One can see from Fig.~\ref{SpinAfig} that
the structure of the measured analyzing power changes in a consistent
way as the mass of the target increases.  Indeed, the depth of the
first minimum and the sharp rise between this and the next maximum
becomes more pronounced with target mass to $^{118}$Sn. As the target
mass increases, the magnitude of this minimum and of the following
maximum are greatly suppressed. The absolute value of those minima and
maxima approaches unity for the heaviest nuclei. As noted in the
discussions of individual results, the $g$--folding calculations
predict observed analyzing powers very well and at scattering angles
for which the cross--section data (usually of magnitude $> 1$~mb/sr)
also are well reproduced.

There is a limited set of measurements \cite{Yoso85} of the spin
rotations for the elastic scattering of 65 MeV polarized protons from
nuclei. These are displayed in Fig.~\ref{SpinRfig}. For the lightest
four nuclei presented, the comparison between the $g$--folding model
results and data is quite good. Those results from the $t$--folding
model calculations do not match observation as adequately.  The
differences between the $g$-- and $t$--folding models are far more
pronounced with the scattering from the heavier targets. While the
$g$--folding model results again give reasonable agreement with the
data, the $t$--folding model results fail to reproduce both shapes and
magnitudes. Indeed, that model predicts a maximum at $\sim 20^{\circ}$
in the spin rotation for the scattering from $^{90}$Zr and $^{118}$Sn;
the data indicate a minimum at that angle and such is predicted by the
$g$--folding model. While there are no data for the scattering from
$^{152}$Sm, that difference between the models is also apparent. Also
by this mass, the first minimum has almost disappeared.  In the case
of scattering from $^{208}$Pb, the result obtained from the
$g$--folding model reflects the shape of the data, although it
overestimates the observed magnitude above $30^{\circ}$. A possible
improvement to this result is to use a different set of SP wave
functions, as is discussed below.

\subsection{Effect of varying the oscillator length}

\begin{figure}
\centering\epsfig{file=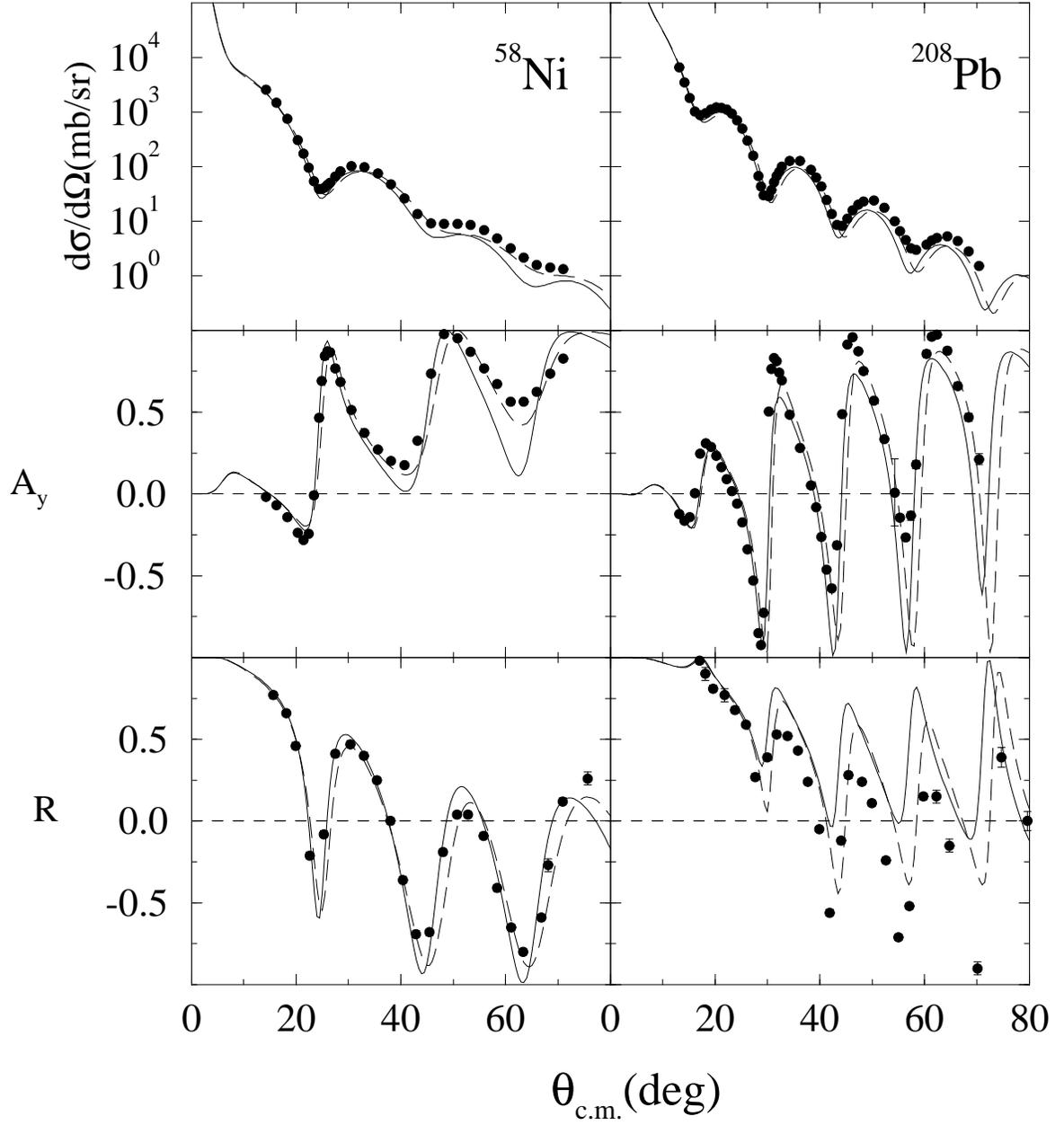,width=\linewidth}
\caption[*]{The differential cross sections (top), analyzing
powers (middle) and spin rotation (bottom)
from the elastic scattering of 65 MeV protons
from $^{58}$Ni and $^{208}$Pb. The data (dots) are
compared with the results of our microscopic model calculations
when differing oscillator lengths for the bound state
wave functions are used. The solid lines represent the 
$b=A^{1/6}$ choice displayed previously (1.97 fm for ${58}$Ni, 
2.43 fm for $^{208}$Pb), while the dashed curves represent fitted
values for $b$ (1.87 fm for ${58}$Ni, 
2.35 fm for $^{208}$Pb).
Data were taken from Refs.~\cite{Yoso85,Saka82}.}
\label{HOfig}
\end{figure}

For the descriptions of the scattering for all nuclei up to and
including $^{20}$Ne we have used SP wave functions which have been set
by fits to electron scattering data. For heavier targets, SP wave
functions of HO form with $b = A^{1/6}$ fm have been used. In general,
this choice has produced very good results in comparison with data,
but it is instructive to investigate the sensitivity of our
calculations to variations in that choice. For this example, we
compare the results of two $g$--folding calculations with scattering
data from $^{58}$Ni and $^{208}$Pb in Fig.~\ref{HOfig}. Therein, the
solid line shows the results of our calculations made using the
standard value. That is 1.97 fm and 2.43 fm for $^{58}$Ni and
$^{208}$Pb, respectively. The dashed lines display the results
obtained when the oscillator length was chosen to give a much better
fit to the cross--section data. Note that these choices are predicated
on the simple structure models assumed for both nuclei.  Specific
shell effects are expected to have some effect on the predictions of
cross sections.  The revised parameters are 1.87 fm and 2.35 fm for
$^{58}$Ni and $^{208}$Pb, respectively. Those lengths were used for
all orbits. The improvement in the results with the new SP wave
functions is very evident. There is much better agreement between the
calculation and experiment for both nuclei, although the position of
the minima in the cross section for the scattering from $^{58}$Ni are
greater than observation. Using these new sets of wave functions, one
observes now much better agreement with the data for both spin
observables. In the case of $^{208}$Pb, the significant improvement in
the spin rotation has been achieved with only a 3\% decrease in the
oscillator length.

\section{Conclusions}
\label{sec_conc}
Optical potentials for the elastic scattering of 65 MeV protons from
nuclei have been obtained by folding medium dependent effective $NN$
interactions with a specification of the ground state for each
nucleus, and also with SP wave functions of either WS or HO
form. Those optical potentials are complex and nonlocal and the
scattering phase shifts and $S$--matrices from which predictions of
the measured quantities were obtained, result by solving the relevant
nonlocal Schr\"odinger equations.

We have obtained good to excellent agreement with scattering data from
targets ranging from $^6$Li to $^{238}$U using the optical potentials
so defined. The framework by which the results were obtained is
entirely parameter-free; no adjustment of any part of the input was
necessary. Thus the nonrelativistic mean-field theory for $pA$
scattering based upon the infinite matter $g$ matrices is reliable for
proton energies down to 65 MeV. This gives encouragement for these
techniques to be used in analyses of data from radioactive beam
experiments. As the optical potentials are derived directly from the
nucleon distributions, instead of averaged charge or matter
distributions, this would provide detailed information on the
structures of exotic nuclei, as was the case in the study of
$^{6,8}$He~\cite{Varenna} and of $^{9,11}$Li~\cite{Li911ref}.

\acknowledgements {This research was supported by a research grant
from the Australian Research Council. We would like to thank
Prof. H. Sakaguchi for providing tabulations of much of the data we
display herein.}


\end{document}